\documentclass[referee]{gji}
\usepackage{timet,color}
\usepackage{enumitem}
\usepackage{amsmath}
\usepackage{mathtools}
\usepackage{nicefrac}
\usepackage{cancel}
\usepackage[normalem]{ulem}
\usepackage{graphics}
\usepackage{algorithm}
\usepackage{algpseudocode}
\usepackage{float}
\usepackage{textcomp}
\usepackage{multirow}
\usepackage{anyfontsize}
\newcommand\upi{\mathord{\mathrm{I}}}
\usepackage{mathtools,cuted}
\usepackage[colorlinks=true,urlcolor=blue,citecolor=blue,linkcolor=blue]{hyperref}
\usepackage[margins]{trackchanges}
\usepackage{csquotes}
\usepackage[pagewise]{lineno}
\usepackage{caption}
\usepackage[position=top]{subcaption}

\usepackage[normalem]{ulem}
\title[Thermomechanical modelling]
{Volumetric (dilatant) plasticity in geodynamic models and implications on thermal dissipation and strain localization}
\author[Momoh, Bhat, Tait, Gerbault]
  {Ekeabino Momoh${^{1*}}$ Harsha S. Bhat$^2$, Stephen Tait$^{1,3}$ and Muriel Gerbault$^{1}$\\
  $^1$  Geosciences Environnement Toulouse - GET, Universit\'{e} Toulouse III - Paul Sabatier,\\  CNRS UMR 5563, IRD, CNES, \emph{31400} Toulouse, France\\
   $^2$ Laboratoire de Geologie, \'{E}cole Normale Sup\'{e}rieure, CNRS UMR 8538, PSL Research University, \\ Paris, \emph{75005} Paris, France \\
$^3$  Laboratoire Dynamiques des Fluides G\'{e}ologiques, Institut de Physique du Globe de Paris,\\ UMR 7154-CNRS, Universit\'{e} de Paris,  \emph{75005} Paris, France\\ 
*Email: {ekeabino.momoh@get.omp.eu,ekemomoh@gmail.com}\\}
\date{\today}

\begin{document}
\label{firstpage}
\maketitle
\begin{summary}
{Here, we present a new thermomechanical geodynamic, numerical implementation that incorporates Maxwell viscoelastic rheology accounting for temperature-dependent power-law dislocation creep and pressure-sensitive, non-associated Drucker-Prager brittle failure, as well as for volumetric stresses and strains during viscoplastic flow, a departure from the traditional incompressible assumptions. In solving for energy conservation, we incorporate the heat source term resulting from irreversible mechanical deformations, which embodies viscoelastic and viscoplastic work, and by considering} the total stress tensor and total inelastic strain rate tensors, including dilatant plasticity effects for lithospheric-scale applications, instead of only the shear terms as is usually assumed for incompressible materials. This form of the work term thus allows to consider, volumetric deformation and to couple the energy equation to the constitutive description, and hence the stress balance, via the evolving temperature field.  Code design enables us to switch individual features of this general rheology \enquote{on or off} and thus to benchmark this implementation with published numerical experiments of crustal-scale shortening experiments.  We investigate whether \enquote{brittle-plastic} compressibility can promote or inhibit localization of deformation and thermal evolution during compression for crustal, and upper mantle rheology.  For both crustal-scale and lithospheric-scale experiments, we establish that the feedback from volumetric dissipation, while contributing to temperature increase along with shear dissipation, can potentially slow down heat production per unit time, depending on the choice of boundary conditions. Our new implementation can be used to address buckling problems and collision tectonics. 
\end{summary}


\section{Introduction}
Strain localization, i.e. the concentration of internal deformation along narrow paths of finite thickness in a material during loading, is a key ingredient of plate tectonics \cite{jacquey2021}. It is typically interpreted as a precursor to material failure \cite{besson2010} and characterizes fault zones, ranging in length from a few meters to hundreds of kilometers and up to plate boundary scale, whose activity may persist for millions of years. In the context of plate convergence in geodynamics, both initiation of subduction interfaces and/or lithospheric-scale structures in continental collision zones are example cases where strain localization and the evolution of deformation are central problems to study \cite{toth1998,lieb1998,stern2004,stern2018,lallemand2021}. An underlying issue is that experimentally determined strengths of rocks (primarily mantle peridotites) are too great for the resistance of a homogeneous lithosphere to be overcome by the stresses generated by buoyant mantle convection, making it difficult to break  \cite{mckenzie1977,cloetingh1989,mueller91}. Strain localization in the lithosphere is associated to an irreversible process and to the formation of shear bands, which may be  compared to real fault zones. The onset of localized deformation in numerical or analogue models is usually controlled either by including some heterogeneity in initial conditions, for example in material properties, and/or some rheologic characteristic that will lead to positive feedback. Thus, the existence of prescribed weak zones such as lithospheric-scale faults inherited from previous tectonic episodes is expected to be relevant \cite{cloetingh1982,gurnis1992,toth1998,nikolaeva2010,baes11}. Rheologic features prone to causing a positive feedback once localization nucleates such as grain-size reduction \cite{thielmann2015}, structural softening \cite{llp13,duretz2016}, thermally-activated softening due to shear heating during irreversible deformation \cite{cramerikaus,thielmann2012,duretz2015,willis2019}, density changes or age offsets between plates  \cite{leng2011,leng2015,zhang2021,zhou2021} material anisotropy \cite{pardoen15} or damage rheology \cite{karrech2011} have been looked at. Thielmann and Kaus \shortcite{thielmann2012}, for example, investigated thermal softening due to shear heating during irreversible deformation and explicitly did not include a pre-existing mechanically weak zone to focus on this thermally-activated feedback. However, their system was given an initial thermal heterogeneity with an age offset to facilitate nucleation of localization. Some of these various effects might indeed occur concomitantly in natural situations. We note here that strain localization can also arise naturally from non-associative brittle plasticity \cite{vermeer1984,gerbault1998}, as a result of kinematic stress rotation effects.

Geodynamic modelling has sometimes been approached from a fluid-mechanical perspective and assumptions of incompressibility, either elastic or plastic \cite{gerya2004,babeyko2008,thielmann2012,schmalholz2014kinematics,ruh2015,vogt2017}, which means that elastic compressibility and/or dilation during plastic flow are unaccounted for. In studies of lithospheric-scale deformation, plastic compressibility is mostly not taken into account \cite{kaus2010,duretz2014,bessat2020}, and elasticity is often neglected in studies over long time scales \cite{garel2014,holt2015,Patocka2019,pajang2021topographic}. However, a number of studies do account for elastic and plastic compressibilities into the constitutive laws for large-scale lithospheric models \cite{hassani1997,gerbault2000,gurnis2004,burov2010,burov2014,duretz2020,duretz2021,jacquey}. From a solid mechanics perspective, rocks are known to accommodate elastic behaviour and undergo volumetric strain during plastic deformation and they are therefore not strictly plastically-incompressible \cite{bridgman1964,cook1970,alonso05,zhao2010}.  At least in the colder parts of the lithosphere, the role of elasticity in stress storage and hence its potential influence on the overall deformational behaviour might actually have been underestimated. 

Whether or not elasticity or plasticity are included in constitutive laws for geodynamic modelling, the influence of volumetric (dilatant) plastic deformation as an energy source in the energy equation has not been considered yet by the geodynamic community. It remains an open question whether or not the plastic compressibility of rocks under lithospheric conditions may contribute to localization both from a solid-mechanical and a thermal perspective.

In the current contribution, we present and benchmark a new solid-mechanical numerical implementation that explicitly includes volumetric (dilatant plastic) contributions in the momentum and energy balances. We include specifically a source term in the energy equation that is due to irreversible deformational work. We explore the effect of thermal softening as an agent of localization, but in addition to considering the impact of heating due to irreversible shear deformation, we aim to develop an understanding of the role of volumetric (dilatant plastic) deformation. In practical terms, we compare the effect of including volumetric plastic deformations in the thermal feedback. We compute irreversible deformation and heating with the full stress and stain-rate tensors, or with only shear terms contributing thermodynamically as has been mostly done before \cite{lieb2001,kaus2006,nikolaeva2010,thielmann2015}, i.e., we can switch on or off the different effects, so that we go beyond previous work. For convenience and to avoid confusion with general volumetric strains associated with thermoelastic effects, and finite deformations, we refer to heating arising from contributions due to irreversible dilatant plastic strains and volumetric stresses as dilatant heating.

We begin by stating the mechanical and energy conservation laws and then give a step-by-step description of the constitutive laws and solution methods we have developed. We then proceed to benchmark our new implementation against published models by presenting comparative tests assuming viscoelastic, and viscoelastic-viscoplastic rheologies, and  compare the relative contributions of shear heating and shear combined with dilatant heating. Furthermore, we apply our approach to the geodynamic context of lithospheric-scale strain localization. Finally, we discuss the implications of this study for future work.

\section{Theoretical framework}
\subsection{Conservation laws}
Our starting point is the balance of linear momentum for a compressible viscoelastic-viscoplastic solid medium given by \cite{lemaitre1994,zienctay}:
\begin{equation}
\dfrac{\partial{{\sigma}_{ij}}}{\partial{x}_j} +f_i=\rho\dfrac{\partial^2{u_i}}{\partial{t}^2},
\label{conservation_equations}
\end{equation}
\noindent where ${\sigma_{ij}}$ represents components of the Cauchy stress tensor, ${f_i}$ represent internal body forces: in this case the lithostatic stress state due to the weight of the material given by ${\rho g}$, with ${\rho}$ as density, ${g}$ as gravitational acceleration, ${x_j}$  represents spatial variables in the Cartesian coordinates; ${{u_i}}$ are the components of displacement. A quasi-static approach is used, which means that we account for small accelerations in the system. The mass balance in Lagrangian framework is given by:
\begin{equation}	\rho_\mathrm{R}=\mathrm{det}\left({\delta_{ij}+\dfrac{\partial u_i}{\partial x_j}}\right)\rho,
\end{equation}
where ${\rho_\mathrm{R}}$ is the mass density at a point in a reference (undeformed) configuration and ${\rho}$ is the density at that point in the current (deformed) configuration. The quantity in parenthesis maps a quantity from a reference (undeformed) state to a current configuration \cite{zienctay}. Under infinitesimal strain theory (where displacements and strains are infinitesimal), mass balance translates to fixed density.
Mechanical dissipation arises from irreversible thermomechanical work, i.e., due to viscoelastic and viscoplastic deformation. We can account for heat generation and diffusion in the conservation of energy which reads:
\begin{equation} 
\dfrac{\partial{T}}{\partial{t}}=\alpha_\textrm{th}{\nabla^2 T}+\beta\dfrac{{{\sigma_{ij}}(\dot\varepsilon^\textrm{v}_{ij}+\dot\varepsilon^\textrm{vp}_{ij})}}{\rho C_p},
\label{eqn:thermal}
\end{equation} 
where ${C_p\;\mathrm{(J\;kg^{-1}\;K^{-1})}}$ is the specific heat capacity and ${\alpha_\textrm{th}\;\mathrm{(m^{2}s^{-1}})}$ is the thermal diffusivity with ${\alpha_{th}=\lambda/\rho C_p}$; ${\lambda}$ being thermal conductivity, which may or may not be constant. ${{{\dot\varepsilon}}^{\textrm{v}}_{ij}}$ and ${{{\dot\varepsilon}}^{\textrm{vp}}_{ij}}$ strain rate components from  (viscous) creep and viscoplastic deformations, respectively; $\sigma_{ij}\left(\dot{\varepsilon}^\textrm{v}_{ij}+\dot{\varepsilon}^\textrm{vp}_{ij}\right)$ is the contribution to heating from deformational work \cite{rittel1999,ravi} with ${\beta}$, the Taylor-Quinney coefficient which quantifies the proportion of deformational work which is dissipated into heat. In this paper, we choose a value of 1, i.e., all mechanical work is dissipated as heat. Which deformation strain rate contributes to the thermal dissipation depends on which of them is dominant at a given point in space or time and is not imposed a priori. In addition, thermoelastic effects including thermal expansion and adiabatic heating, and density variation with temperature and pressure are not considered here, since these variations are assumed infinitesimal under small strain theory. 
\subsection{Deformational framework}\label{deformedframework}
Our formulation relies on a small (infinitesimal) strain theory which allows an additive decomposition of any contributing strain rate components \cite{mase1970,lemaitre1994,zienctay,bower,neto,deborst}. Our constitutive law incorporates elastic strain, ductile creep, and plastic flow laws with inelastic work being dissipated into heat.  

We assume a Maxwell viscoelastic rheology and that the strain rate is a combination of an elastic contribution and a viscous contribution \cite{turcotte2002,deborst,geryabook}, and Drucker-Prager viscoplasticity. Each time step starts with an elastic trial state that is then corrected via viscous creep activated at non-zero stresses, and followed by plasticity through a yield stress criterion. Such rheology has been widely used in earlier geodynamic models \cite{hassani1997,gerbault2000,gerbault2002,burovclo2010}. The standard approach for implementing viscoelasticity in the geodynamic community is a combination of shear modulus ${G}$, viscosity ${\eta}$ and time step ${\Delta t}$. Recent considerations involve a correction of the shear modulus to obtain a viscoelastic shear modulus (${G^\textrm{ve}}$) that depends on the dynamic shear viscosity (${\eta}$), the elastic shear modulus (${G}$) and the algorithmic time step (${\Delta t}$), \cite{duretz2018,duretz2021}. 

Based on small-strain assumption, we compute the stress evolution, creep and plastic flow relying on an additive decomposition of the total strain rate tensor into elastic, viscous, and viscoplastic strain rates, represented, respectively, by superscripts \enquote{$\textrm{e}$}, \enquote{$\textrm{v}$} and \enquote{${\textrm{vp}}$} below:
\begin{equation}
{\dot\varepsilon}_{ij} = {\dot\varepsilon}^\textrm{e}_{ij}+{\dot\varepsilon}^\textrm{v}_{ij}+{\dot\varepsilon}^\textrm{vp}_{ij},
\label{rateform}
\end{equation}
\begin{equation}
{\dot\varepsilon}_{ij} = {\dot\varepsilon}^\textrm{e}_{ij}+{\dot\gamma^\textrm{v}}\dfrac{\partial\Phi^\textrm{v}_\textrm{F}}{\partial s_{ij}}+{\dot\gamma^\textrm{vp}}\dfrac{\partial\Phi^\textrm{vp}_\textrm{F}}{\partial\sigma_{ij}}.
\label{decomposition0}
\end{equation}
${\Phi^\textrm{v}_\textrm{F}}$ and ${\Phi^\textrm{vp}_\textrm{F}}$ are respectively viscous (creep) and viscoplastic flow potentials, the derivatives of the viscous and viscoplastic flow potentials with respect to deviatoric stress and total stress components are indicative of viscous and viscoplastic flow directions, respectively; ${{\dot\gamma}^\textrm{v}}$ and ${{\dot\gamma}^\textrm{vp}}$ are viscous and viscoplastic multipliers in rate form; ${s_{ij}}$ represent the components of the deviatoric stress tensor with ${{s_{ij}=\sigma_{ij}}-\sigma_{kk}/3}$. 

\subsubsection{Elastic Rheology}
The elastic part of the material response is treated as an isotropic solid characterized by its Young's modulus, $E$, and Poisson's ratio, $\nu$. The classical elastic stress strain relation is given by Hooke's Law \cite{zienctay,bower}:
\begin{equation}
{\sigma}_{ij}^\textrm{e}=C_{ijkl}^\textrm{e}{\varepsilon}_{kl}^\textrm{e},
\end{equation} 
where ${C_{ijkl}^\textrm{e}}$ is a fourth-order tensor representing the components of the elastic stiffness tensor of a material, incorporating the elastic constants listed above. ${{\sigma}_{ij}^\textrm{e}}$ and ${\varepsilon}^\textrm{e}_{kl}$ are respectively, components of elastic stresses and strain tensors. 

\subsubsection{Creep (Viscous) Rheology}

It is commonly assumed that, at high temperatures and low stresses, rocks deform viscously without stresses necessarily overcoming a yield criterion, i.e., yield criterion is effectively zero \cite{neto}. The limiting behaviour for the elastic rheology in our constitutive law is therefore high temperature creep. We assume that the dominant creep mechanism is dislocation creep and formulate the viscous deformation as follows: 
\begin{equation}
\dot{\varepsilon}_{ij}^\textrm{v} = \dot{\gamma^\textrm{v}}({\sigma_{ij}^\textrm{e}},T)\dfrac{\partial{\Phi_\textrm{F}^\textrm{v}}}{\partial{s}_{ij}^\textrm{e}},
\label{viscousflow}
\end{equation}
with ${\dot{\gamma^\textrm{v}}}$, a non-negative quantity specifying a magnitude of viscous flow in rate form and ${{\partial{\Phi_\textrm{F}^\textrm{v}}}/{\partial{s}_{ij}^\textrm{e}}}$ specifying the direction of viscous flow taking account of only deviatoric stresses. $\Phi_\textrm{F}^\textrm{v} = \sqrt{J_\textrm{II}^\textrm{e}}$ with ${{J_\textrm{II}^\textrm{e}}}$ representing the invariant of the deviatoric stress tensor,  is the viscous flow potential whose direction is given by: 
\begin{equation}
 \dfrac{\partial{\Phi_\textrm{F}^\textrm{v}}}{\partial{s}_{ij}^\textrm{e}}=\dfrac{{s}_{ij}^\textrm{e}}{2\sqrt{J_\textrm{II}^\textrm{e}(s_{ij}^\textrm{e})}}.
\end{equation}

The magnitude of viscous flow $\dot{\gamma^\textrm{v}}$ assumes various functional forms depending on the specific problem and material \cite{neto}. We utilize the power-law functional form of \cite{boyle,rosakis2000,poulet2016}:
\begin{equation}
\dot{\gamma}^\textrm{v}({\sigma_{ij}^\textrm{e}},T)=\left({\Phi^\textrm{v}_\textrm{F}}\right)^mf(T).
\label{functional_form}
\end{equation}
$m$ represents the power law exponent that describes the sensitivity to stress during viscous flow. The temperature dependence is given in the form of \cite{skrzypek,kohlstedt,ranalli}: 
\begin{equation}
f(T)=A{\textrm e}^{-\frac{E_a}{RT}}.
\label{disloc}
\end{equation}
Here $E_a$ is the activation energy, $R$ is the molecular gas constant, $T$ is the absolute temperature (in Kelvin, K) and $A$ is the pre-exponential factor.

Equation \ref{disloc} indicates that viscous creep becomes significant at high temperatures; otherwise, the viscous correction at low temperatures is so small that the effect of elasticity is preserved. Because the viscous correction term is insignificant at low temperatures and low stresses, the feedback to viscous strain rate and deviatoric stresses is small, and the contribution to thermal dissipation evolves similarly (thermoelastic effects are unaccounted for). Therefore, the partitioning between elastic behaviour and ductile creep depends essentially on the temperature. This strategy can be seen as an elastic predictor, followed by a viscous correction step \cite{jacquey}.

\subsubsection{Viscoplastic Rheology} 

The elastic and viscous behaviour or viscoelastic rheology is bounded by a pressure-sensitive frictional plastic yield criterion which has been used to model brittle deformation on a lithospheric scale \cite{moresi07,babeyko2008,burovclo2010,kaus2010,baes11,jacquey}. Here, we use a Drucker-Prager criterion given by \cite{dpg1952,alej}:
\begin{equation} 
\Phi_\textrm{Y}^\textrm{DP}= \sqrt{J_\textrm{II}^\textrm{v}(s_{ij}^\textrm{v})}+\alpha_1P^\textrm{e}-\alpha_2c > 0.
\label{drucker}
\end{equation}
\noindent Where $J_\textrm{II}^\textrm{v}=s_{ij}^\textrm{v}s_{ij}^\textrm{v}/2$ represents the second invariant of the deviatoric stress tensor after creep deformation, ${P^\textrm{e}=I_1/3=\sigma_{kk}/3}$ is the first invariant of the stress tensor, ${c}$ is the cohesion which may depend on the deformation history; ${\alpha_1}$ and ${\alpha_2}$ are material-dependent constants which are functions of the internal friction angle (${\varphi}$) as follows for the specific case of a plane strain deformation \cite{neto}:
\begin{equation} 
\alpha_1=\dfrac{3 \tan{\varphi}}{\sqrt{9+12\tan{^2\varphi}}}, \alpha_2=\dfrac{3}{\sqrt{9+12\tan{^2\varphi}}}.
\end{equation}

Viscoplastic flow rate initiates when rate-dependent plastic deformation has begun and follows the standard formulation \cite{desai1987,vermeer1990,alrub}:
\begin{equation}
\dot{\varepsilon}_{ij}^\textrm{vp} = \dot{\gamma}^\textrm{vp}({\sigma_{ij}^\textrm{v}},T)\dfrac{\partial{\Phi_\textrm{F}^\textrm{vp}}}{\partial{\sigma}_{ij}^\textrm{v}} =  \dot{\gamma}^\textrm{vp}({\sigma_{ij}^\textrm{v}},T)\dfrac{\partial}{\partial{\sigma}_{ij}^\textrm{v}}\left(\sqrt{J_\textrm{II}} + \alpha_3 P\right) 
\label{viscoflow}
\end{equation}
specifying a magnitude and direction of viscoplastic flow, with ${\dot{\gamma}^\textrm{vp}}$ as a non-negative quantity defined as the viscoplastic consistency parameter describing the magnitude of viscoplastic flow \cite{simo1998}; $\Phi_\textrm{F}^\textrm{vp}$ is the viscoplastic flow potential and $\alpha_3=3 \tan{\psi}/ \sqrt{9+12\tan{^2}\psi}$, ${\psi}$ is the dilatancy angle whose value may describe associative or non-associative plasticity \cite{vermeer1984}. The derivative of the viscoplastic flow potential with respect to the stress tensor describes the direction of viscoplastic flow: 
\begin{equation}
 \dfrac{\partial{\Phi_\textrm{F}^\textrm{vp}}}{\partial{\sigma}_{ij}^\textrm{v}}=\dfrac{{s}_{ij}^\textrm{v}}{2\sqrt{J_\textrm{II}^\textrm{v}(s_{ij}^\textrm{v})}}+ \dfrac{\alpha_3}{3}\delta_{ij}.
\end{equation}
The equation above accounts for both deviatoric and dilatant viscoplastic flow. Non-zero dilatancy removes the assumption of a plastically incompressible material, i.e., one for which pressure changes due to viscoplastic deformation do not result in a net volume change in the material \cite{poliakov1994,poliakov1994a,turcotte2002}. In this crucial respect, our work differs from widespread practice in geodynamics, which approximates rocks as either elastically or plastically incompressible \cite{babeyko2008,schmeling2008benchmark,kaus2010,leng2015,ruh2015,jaquet2016,kiss2020,pajang2021topographic}.

The viscoplastic consistency parameter ${\dot\gamma^\textrm{vp}}$ can represent various functional forms depending on the problem and material \cite{neto}, we utilize the functional form of \cite{zien1974,owen80,perzyna1986,desai1987}:
\begin{equation}
{\dot{\gamma}^\textrm{vp}({\sigma_{ij}^\textrm{v}},T)=\dfrac{1}{\mu}\left<{\dfrac{\Phi_\textrm{Y}^\textrm{DP}}{\Phi_{0}}}\right>^m}
\label{functionals_vp}
\end{equation}
where ${\left<\cdot\right>}$ is the Macaulay bracket defined for any function ${f}$ as: ${\left<f\right> = f\; \mathrm{if}\;f> 0}$ and ${\left<f\right> = 0\; \mathrm{if}\;f\leq 0}$. $m$ here represents a material parameter following the form of Desai and Zhang \shortcite{desai1987}. ${\Phi_\textrm{Y}^\textrm{DP}}$  represents the Drucker-Prager yield criterion, while ${\Phi_\textrm{0}}$ is a normalizing term usually taken as the yield stress or cohesion \cite{desai1987}, or taken as the plastic viscosity \cite{jacquey}. 
Considering viscoplasticity, the model requires an additional term, ${\mu}$ in Equation \ref{functionals_vp}, where ${1/\mu}$ has units of the inverse of time and expresses the relative rate of viscoplastic strain \cite{desai1987}. The temperature dependence of plastic deformation is included in the yield function ${\Phi_\textrm{Y}^\textrm{DP}}$ through ${s_{ij}^\textrm{v}}$. In principle, we can include frictional hardening (or softening) and cohesion hardening (or softening) depending on the deformation history through a variety of strategies, \cite{leroy1989,ortiz1990}, but these will not be considered here for the sake of simplicity.

\subsubsection{Sources of volumetric strains} 

Volumetric strains measure volume changes in a material \cite{bower}. They are often suppressed in geodynamic codes which assume incompressibility by setting the divergence of velocity to zero \cite{kaus2010,thielmann2012,duretz2015,jourdon2018,bessat2020}. Since the volumetric strains are related to the pressure through the bulk modulus, setting it to an arbitrarily high value can suppress volumetric elastic strains \cite{zienctay}. These are volume changes from elastic rheologies. In terms of plasticity, not all flow laws accommodate dilatant behaviour, for example, the von Mises criterion only utilizes the deviatoric stress invariant and cohesion to define the plastic strength of a material; other flow laws can suppress the influence of dilatant strain when the dilatancy angle is set to zero \cite{babeyko2008}, thereby suppressing volume changes due to plastic flow.

In addition to accounting for elastic compressibility in our constitutive laws, we include contributions to volumetric strains from dilatant plasticity during irreversible brittle deformation. Our point will be to show specifically the impact of  this brittle plastic compressibility (dilatant plastic strain)  on the deformation state and thermal feedback.

\subsection{Numerical implementation}

For computational purposes, we express Equations \ref{rateform} and \ref{decomposition0} in incremental form:
\begin{equation}
{\Delta\varepsilon}_{ij} = {\Delta\varepsilon}^\textrm{e}_{ij}+{\Delta\varepsilon}^\textrm{v}_{ij}+{\Delta\varepsilon}^\textrm{vp}_{ij},
\label{incrmentalform}
\end{equation}
\begin{equation}
{\Delta\varepsilon}_{ij} = {\Delta\varepsilon}^\textrm{e}_{ij}+{\Delta\gamma^\textrm{v}}\dfrac{\partial\Phi^\textrm{v}_\textrm{F}}{\partial s_{ij}}+{\Delta\gamma^\textrm{vp}}\dfrac{\partial\Phi^\textrm{vp}_\textrm{F}}{\partial\sigma_{ij}}.
\label{decomposition1}
\end{equation}

Similarly, the unknowns in Equations \ref{functional_form} and \ref{functionals_vp} which were expressed in rate form are expressed in incremental form.

In solving the conservation equations, we have utilized a solid mechanical finite element solver, Abaqus\textsuperscript{\textcopyright} \cite{abaqustheory,abaqus}. Abaqus\textsuperscript{\textcopyright} is a robust, optimized, engineering industry-accepted finite element solver with a wide array of rheologies and element library. The software package also allows a user to custom any rheology or process through subroutines. Whenever stresses are updated through customized subroutines, a Consistent Algorithmic Tangent Modulus is required to aid convergence and estimate a time increment for the next simulation time step. Abaqus\textsuperscript{\textcopyright} has been widely used to study a variety of geodynamic problems \cite{lieb1998,branlund2000,branlund2001,lieb2001,lieb2003,gerbault2002,dyksterhuis2005,capitanio2007dynamic,salomon2018}. We have used Abaqus\textsuperscript{\textcopyright} capabilities to solve the mechanical conservation equations, while we implemented the constitutive laws and the energy conservation problem through our own customized subroutines. We therefore transcribed our mechanical and thermal constitutive laws described in Section \ref{deformedframework} into Fortran subroutines: a User-Material (UMAT) and User-Material Heat Transfer (UMATHT), respectively, implementable in Abaqus\textsuperscript{\textcopyright}.

The details of the algorithmic implementation of our constitutive laws, including the detailed computation of the Consistent Algorithmic Tangent Moduli for viscoelastic and viscoplastic rheologies, are given in Appendix \ref{AppendixA}. A summary of the thermomechanical implementation is illustrated in a pseudocode in Algorithm \ref{algorithm1}. Our implementation can be summarized as an elastic trial state, a thermally-activated viscous correction to the elastic state (viscoelastic rheology), followed by a frictional plasticity correction activated upon the material stress state above the Drucker-Prager yield criterion.
\begin{figure}
\begin{algorithm}
{Pseudo-code for custom viscoelastic-viscoplastic constitutive law, over one time-step loop.}
\label{algorithm1}
\begin{algorithmic}
\Require ${\Delta\varepsilon^\textrm{e, trial}_n}$ at time step {n}
\State From ${\Delta\varepsilon^\textrm{e, trial}_\textrm{n} ,\; \textrm{obtain} \; \sigma^\textrm{e, trial}_n}$ 
\If{$\sqrt{J_\textrm{II}^\textrm{e}}\neq 0$} \State Start creep routine
\While  {${\tilde{\Phi}^\textrm{v}>\textrm{Tolerance}}$}\State {Newton-Raphson iterations to obtain ${\Delta\gamma^\textrm{v}}$}
\EndWhile
\State Update ${s_\textrm{n}^\textrm{v}}$ and ${J_\textrm{II}^\textrm{v}(s_\textrm{n}^\textrm{v})}$
\State Compute ${\Phi_\textrm{Y}^\textrm{DP}(s_\textrm{n}^\textrm{v})}$
\If{${\Phi_\textrm{Y}^\textrm{DP} > 0}$} \State Viscoplastic routine
\While  {${\tilde{\Phi}^\textrm{vp}_\textrm{R}>\textrm{Tolerance}}$}\State {Newton-Raphson iterations to obtain ${\Delta\gamma^\textrm{vp}}$}
\EndWhile
\State Viscoplastic updates
\State Update ${\varepsilon_\textrm{n+1}^\textrm{vp},s_\textrm{n+1}^\textrm{vp}, P_\textrm{n+1}, \sigma_\textrm{n+1}, T_\textrm{n+1}}$.  
\State Consistent algorithmic tangent modulus. 
\State Compute {${C_{ijkl}^\textrm{vp}}$} 
\Else
\State Viscous updates
\State Update ${\varepsilon_\textrm{n+1}^\textrm{v},s_\textrm{n+1}^\textrm{v}, \sigma_\textrm{n+1}, T_\textrm{n+1}}$.
\State Consistent algorithmic tangent modulus  
\State Compute {${C_{ijkl}^\textrm{v}}$}  
\EndIf
\Else
\State The stress state is elastic
\State {${\sigma_\textrm{n+1}=\sigma_\textrm{n}^\textrm{e, trial}}$} 
\EndIf
\end{algorithmic}
\end{algorithm}
\end{figure}
At each simulation time step for all integration points (element nodal), Abaqus calls our UMAT subroutine to compute the stresses and uses the stresses to solve the mechanics (Equation \ref{conservation_equations}). The consistent algorithmic tangent modulus is then used to estimate the adaptive simulation time step to be used for the next time increment.

In coupling the mechanical deformations (Equation \ref{conservation_equations}) to the energy conservation problem (Equation \ref{eqn:thermal}), we compute the heat source given by the second term on the right hand side of (Equation \ref{eqn:thermal}, in the UMAT after updating the stresses which is then passed to the User-Material Heat Transfer (UMATHT) Fortran subroutine. Finally, the stresses, temperature, and other state variables are stored, which are called at the next time step as history-dependent variables.

Since we are interested in the feedback from mechanical work to thermal dissipation and vice versa, we utilized the 2-D coupled continuum plane strain temperature-displacement triangular element library (CPE3T) with linear shape functions within the Abaqus\textsuperscript{\textcopyright} element library. The CPE3T element library is composed of 3 nodal points, with two of them having displacement degrees of freedom and one temperature degree of freedom. 

In addition to accounting for dilatant heating terms in the energy budget, another originality of our approach is that we utilize two rounds of Newton-Raphson iterations to compute the viscoelastic and viscoplastic multipliers. To ensure a strict positivity of these terms for the respective rheologies and convergence of the local Newton-Raphson iterations, we include a simple bisection scheme within the Newton-Raphson loops \cite{net,chapra}.

\section{Algorithmic testing and validation}

To assess the performance of our constitutive laws, we investigated shear band formation using viscoelastic rheology and then, viscoelastic-viscoplastic rheology. We first proceeded by investigating the formation of ductile shear bands using viscoelastic rheology as in Duretz et al \shortcite{duretz2014} with an initial constant temperature. Thereafter, we investigated crustal-scale shear band formations in brittle and ductile regimes using viscoelastic-viscoplastic rheology as used by Duretz et al \shortcite{duretz2021}. We utilized the same geometry and boundary conditions for a systematic comparison with published results and highlighted similarities and differences in our different constitutive approaches.

\subsection{TEST 1 (benchmark test 1): localization in an isothermal viscoelastic medium\label{benchmark1}}

\subsubsection{Model configuration and boundary conditions}

\begin{figure}
\centering
\includegraphics[width=8.5cm]{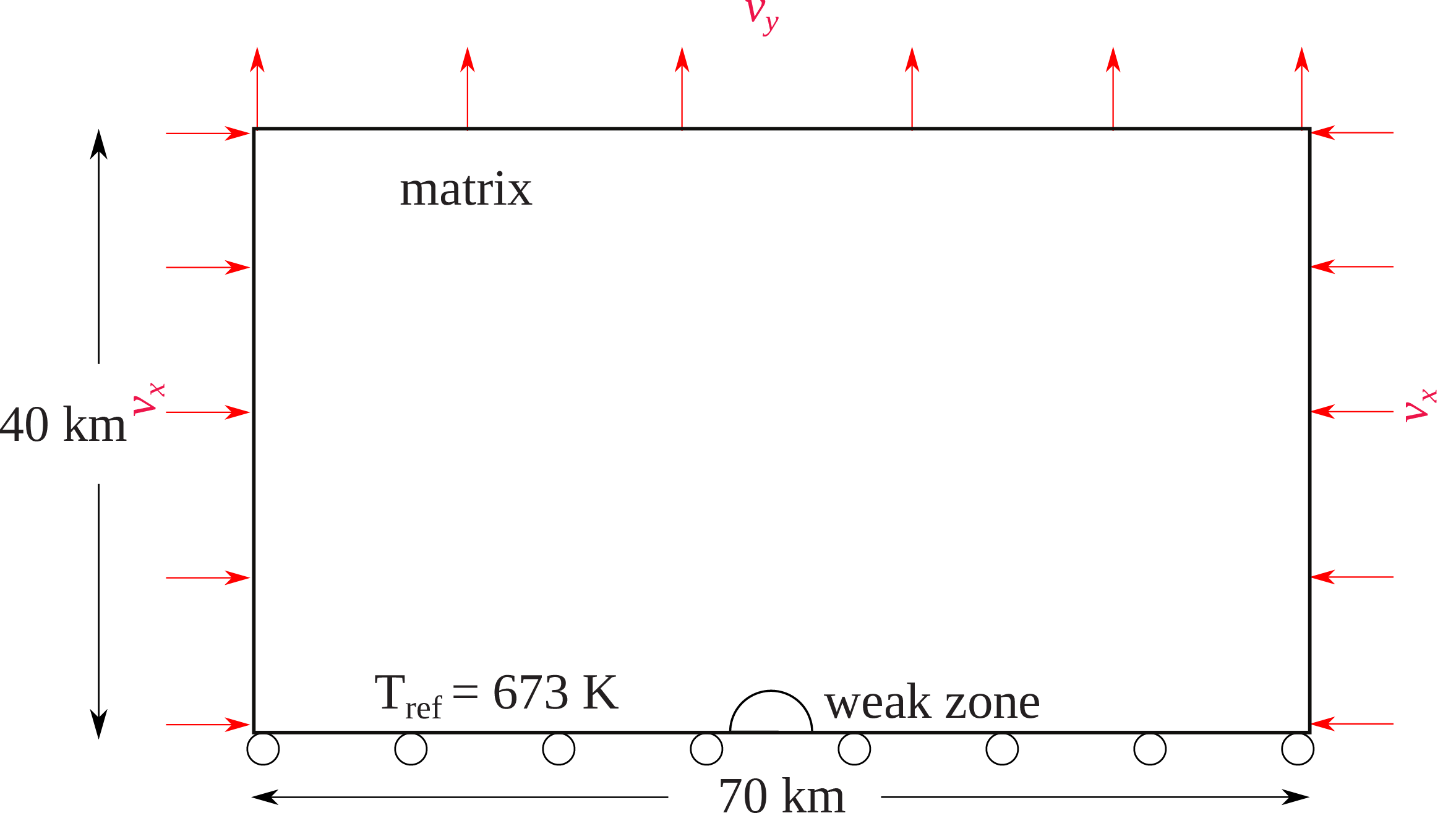}
\caption{TEST 1, input model for studying crustal-scale shear-banding in an isothermal medium with a rock matrix characterized by Maryland diabase rheology and a semi-circular weak zone of radius 3 km characterized by Westerly granite. The setup was introduced by Duretz et al. \protect\shortcite{duretz2014} to study the impact of viscous shear heating on the thickness of shear zones. Velocities are imposed on the top and lateral boundaries to satisfy a background strain rate of ${5\times10^{-14}\textrm{ s}^{-1}}$; all boundaries have zero shear stresses, and the bottom of the model is free-slip.}
\label{duretz14setup}
\end{figure}

Here, we utilized the setup introduced by Duretz et al. \shortcite{duretz2014}. The model setup has dimensions of 70 km by 40 km, comprises a rock matrix that approximates a Maryland diabase rheology and a 3 km radius semi-circular weak inclusion which approximates dry Westerly granite rheology centered on the bottom boundary of the model as shown in Figure \ref{duretz14setup}. The material properties for the matrix and weak inclusion are shown in Table \ref{duretz2014table1}. 

The model boundary conditions approximate a pure shear experiment; therefore we apply velocities on the lateral and top boundaries to satisfy a constant background strain rate of ${5\times10^{-14}}$ s${^{-1}}$, i.e., a time-varying velocity boundary condition; while the bottom boundary is free to slip. All boundaries are thermally insulated so that no heat is evacuated to the surroundings or admitted therefrom. The model was discretized with 20,350 triangular elements connected by 10,477 nodes with linear shape functions, accounting for displacement and temperature degrees of freedom. The element type we use is coupled thermal and displacement elements in plane strain configuration which can help in the efficient conversion of mechanical work to heating provided the heat source is estimated.

As Duretz et al \shortcite{duretz2014} used viscous rheology for an incompressible fluid with viscous feedback to heating (i.e., heat source term described by the inner product of the viscous deviatoric stresses and the viscous strain rate tensors), we utilized the viscoelastic approach with viscous thermal feedback for our test. We utilized a Young's modulus of 25 GPa. We did not enforce a zero velocity divergence condition as is done for incompressible flow \cite{duretz2014}. While the power law rheology is formulated in terms of effective viscosity in \cite{duretz2014}, we implement our creep rheology in terms of strain rate. An additional difference is that we utilized a finite element approach allowing the elements to deform with the model, while Duretz et al. \shortcite{duretz2014} utilized the finite difference/marker in cell approach \cite{gerya2003fd}. Finally, we adapted our time stepping based on the deformation state at the preceding time step, while fixed time steps were used in the reference study \cite{duretz2014}.

{{\begin{table*}
\begin{minipage}{160mm}
\begin{center}
\caption{TEST 1 rheological parameters used in the benchmark setup shown in Figure \ref{duretz14setup} drawn from Duretz et al. \protect\shortcite{duretz2014}.}
\begin{tabular}{@{}llcccccl}
\hline
Material & ${A}$ (Pa${^{-m}}$s${^\textrm{-1})}$ & ${m}$ & $E_a\; $ (kJ mol${^\textrm{-1})}$& ${\alpha_{th}}$ (m${^\textrm{2}}$s${^\textrm{-1}}$) & ${C_p}$ (J kg${^\textrm{-1}}$K${^\textrm{-1}}$) &${\rho}$ (kgm${^\textrm{-3}}$)&\\
\hline
Matrix&3.20${\times}$10${^\textrm{-20}}$&3&276&${8.82\times{10^{-7}}}$ &1050&2700\\
Inclusion&3.16${\times}$10${^\textrm{-26}}$&3.3&186&${8.82\times{10^{-7}}}$ &1050&2700\\
\hline
\end{tabular}
\label{duretz2014table1}
\end{center}
\end{minipage}
\end{table*}}}

\subsubsection{Results of viscoelastic benchmarking}
\begin{figure}
    \centering
\includegraphics[width=12cm]{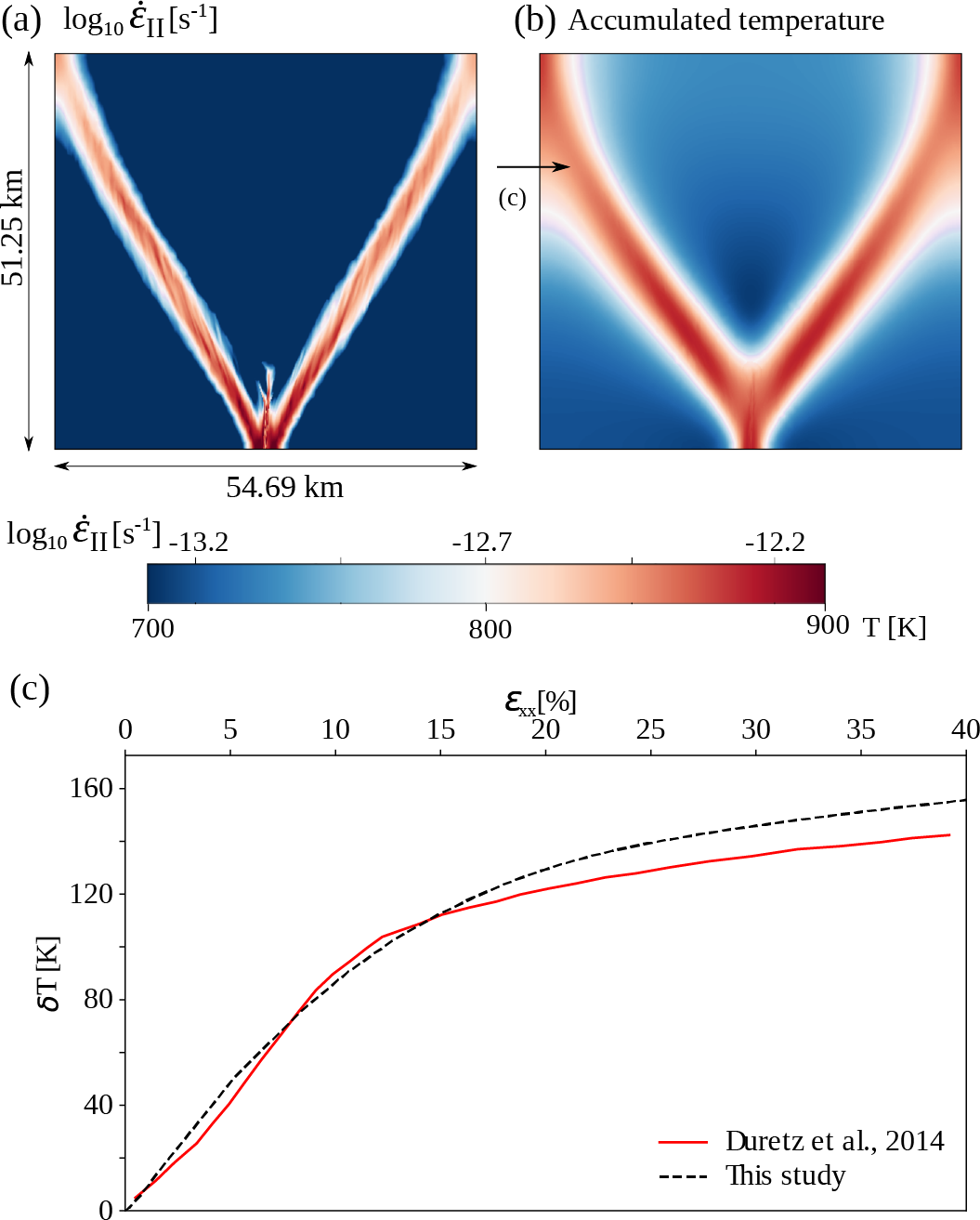}
\caption{TEST 1 results: (a) viscoelastic strain rate, (b) accumulated temperature shown in deformed configuration. The bulk shortening is 25${\%}$ and a background strain rate of 5${\times{10^{-14}\textrm{ s}^\textrm{-1}}}$. (c) Temperature increase for an element within the shear band location of element indicated by an arrow in Figure \ref{duretz14results1}b compared to temperature evolution within the shear zone of Duretz et al. \protect\shortcite{duretz2014}.}
\label{duretz14results1}
\end{figure}

Localized deformation originated from the weak inclusion and propagated symmetrically towards the top left and top right corners of the model domain. In terms of the geometry of the model compared to the benchmark setup \cite{duretz2014}, we were able to realize the same amount of bulk shortening (Figure \ref{duretz14results1}a).
The logarithm of the second invariant of strain rate tensor, ${\sqrt{{\dot\varepsilon_{ij}}{\dot\varepsilon_{ij}}/2}}$, is shown in Figure \ref{duretz14results1}a after an estimated 25${\%}$ of bulk shortening. This evolution of strain rate reproduces the results of the reference model \cite{duretz2014} in terms of the amplitudes and width of the shear bands. We note some smearing as the shear bands propagated towards the boundaries. Shown in Figure \ref{duretz14results1}b is the temperature accumulated at the end of the simulation due to viscous shear heating. This pattern aligns with the viscoelastic deformation, which is the source of the dissipative heating. Temperature rose by up to 205 K within the shear band.

We extracted a 1-D plot of the difference between the final temperature and the initial temperature in an element within the shear band near the top left corner of the deformed model of Figure \ref{duretz14results1}b, and compared the evolution with that of Duretz et al. \shortcite{duretz2014} as shown in Figure \ref{duretz14results1}c. The temperature perturbation is similar to that previously obtained by Duretz et al. \shortcite{duretz2014} until about 15${\%}$ of axial shortening, after which a difference of 5 K builds up at 25${\%}$ of axial shortening. However, this difference of 5 K between our results and the reference study only represents about 3${\%}$ given the large temperature increase. We attribute this difference to differences in the numerical schemes mentioned previously, which is further supported by the supplementary tests carried in the following section.

\subsubsection{Sensitivity to different element types, linear and non-linear interpolators}
\begin{figure}
    \centering
\includegraphics[width=8cm]{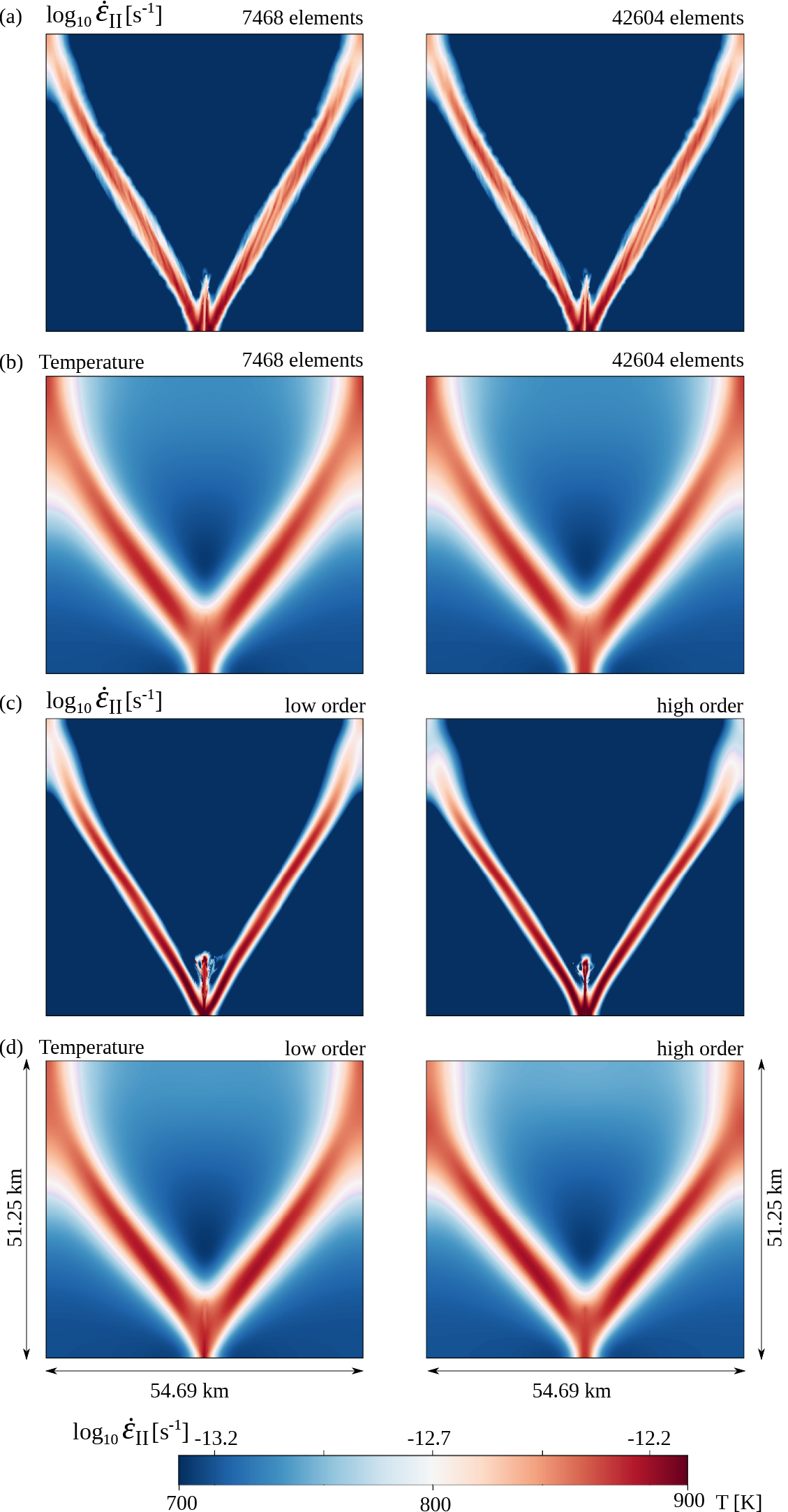}
\caption{TEST 1, resolution tests and element type with linear and non-linear interpolators: (a) logarithm of the second invariant of strain rate for a coarse and finer triangular mesh compared to Figure \ref{duretz14results1}a, (b) corresponding temperature for the coarse and finer meshes, (c) logarithm of second invariant of strain rate invariant using linear (low order) and non-linear (high-order) 20,411 quadrilateral elements, and (d) corresponding temperature for using low-order and high-order elements (linear versus quadratic interpolants). This experiment was done for 25${\%}$ of shortening.}
\label{duretz14resultssensitivity}
\end{figure}

In order to assess the mesh sensitivity of our viscoelastic implementation, especially the influence of mesh types and the type of interpolation used, we used the set up shown in Figure \ref{duretz14setup}. For the results presented in Figure \ref{duretz14results1}a, we used linear triangular elements in which the displacements between nodes were interpolated linearly. In assessing the robustness of our development, we carried out resolution tests using coarser (7,468) and finer (42,604) triangular elements compared to the elements used in Figure \ref{duretz14results1}a. The resolution tests indicate that the shear bands are resolved irrespective of the resolution (Figures \ref{duretz14resultssensitivity}a-b). Using different element types (quadrilateral elements) and including non-linear (quadratic) shape functions, which interpolate displacements between nodes using higher order polynomials, we also found that the deformation was not mesh sensitive (Figure \ref{duretz14resultssensitivity}c-d).

We utilized the same setup as shown in Figure \ref{duretz14setup} but with 20,411 quadrilateral elements, with linear and non-linear shape functions. As shown in Figure \ref{duretz14resultssensitivity}, the dimensions of the deformed domain as well as the width of the major shear bands are essentially identical to those obtained using triangular elements (Figure \ref{duretz14results1}a). The results are, however, insensitive to whether we used low-order or high-order finite elements.  Apart from slightly higher strain rates when quadrilateral elements were used compared to triangular elements, the width of the shear bands and the associated temperature were insensitive to the different mesh types, interpolating functions, and resolution.

\subsection{TEST 2 (benchmark test 2): localization in brittle and ductile regimes}

\subsubsection{Model setup and boundary conditions}
\begin{figure}
\centering
\includegraphics[width=9cm]{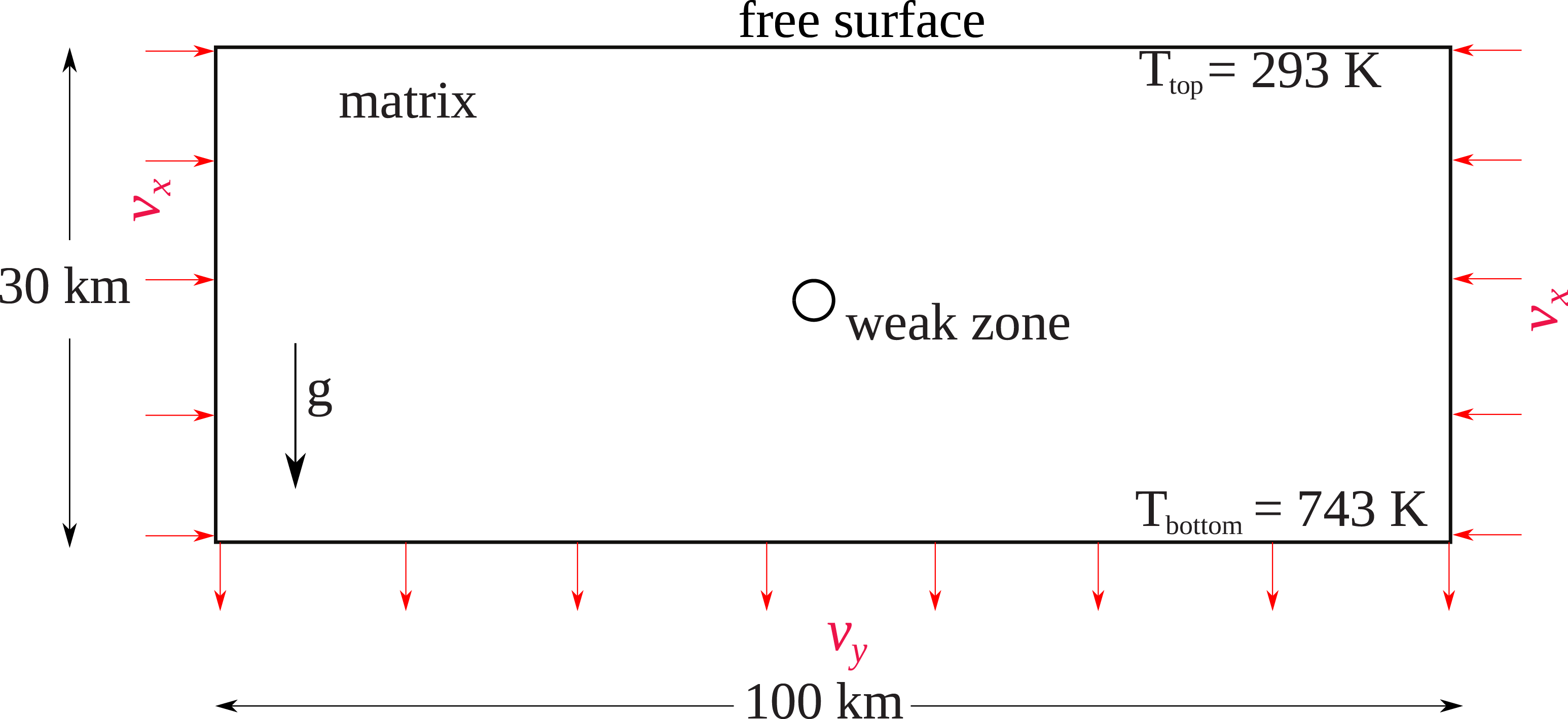}
\caption{TEST 2: input model configuration for studying crustal-scale shear-banding in a pure-shear compression experiment with a rock matrix characterized by Westerly Granite rheology and a circular weak zone of radius 2 km at the center. Velocities are imposed as shown to satisfy a background of ${10^{-15}\textrm{s}^{-1}}.$ The temperature at the top is 293 K with a linear gradient to 739 K at the bottom. We include a pre-loading gravity stage before imposing the velocity boundary conditions. The top surface is kept as a free surface throughout. The model setup and boundary conditions were introduced by Duretz et al. \protect\shortcite{duretz2020,duretz2021}.}
\label{duretz20setup}
\end{figure}

The setup discussed for viscous shear-banding in a ductile regime in Section \ref{benchmark1} above corresponds to an isothermal regime without body forces and with viscoelastic rheology to conform with the reference model \cite{duretz2014}. To test a case incorporating our viscoelastic-viscoplastic rheology, we utilize another setup introduced by Duretz et al. \shortcite{duretz2020,duretz2021}; it assumes a  100 km long and 30 km thick crust with rheology corresponding to that of Westerly granite, and in which a 2 km radius circular imperfection is embedded in the middle of the model, as shown in Figure \ref{duretz20setup}. The model includes an initial temperature of 293 K at the top increasing linearly to 739 K at the base of the model (15 K/km). All boundaries are insulated to prevent heat loss to the surroundings or input from external sources, such that dissipative heating only arises from the internal deformation. The material properties are shown in Table \ref{duretz2020table}. We carried out simulations to keep a constant boundary strain rate of 10\textsuperscript{-15} s\textsuperscript{-1}. Note that we used an additional parameter ${\mu}$ described in Equation \ref{functionals_vp}, where ${1/\mu}$ is the relative rate of viscoplastic strain, chosen to be 10\textsuperscript{-15} s\textsuperscript{-1}. 

Points of departure from the results of Duretz et al. \shortcite{duretz2021} and our model arise from several differences in the numerical methods, which were mentioned already above. Furthermore, in Duretz et al. \shortcite{duretz2018,duretz2021}, the conservation equations were discretized using a staggered grid finite difference scheme where velocity and pressure are considered as primary variables, as traditionally assumed by the geodynamic community; but here, we have discretized our system of equations on a finite element grid with displacement as primary variable, following a typical engineering approach \cite{zienctay}. 

\begin{figure}
    \centering
\includegraphics[width=14cm]{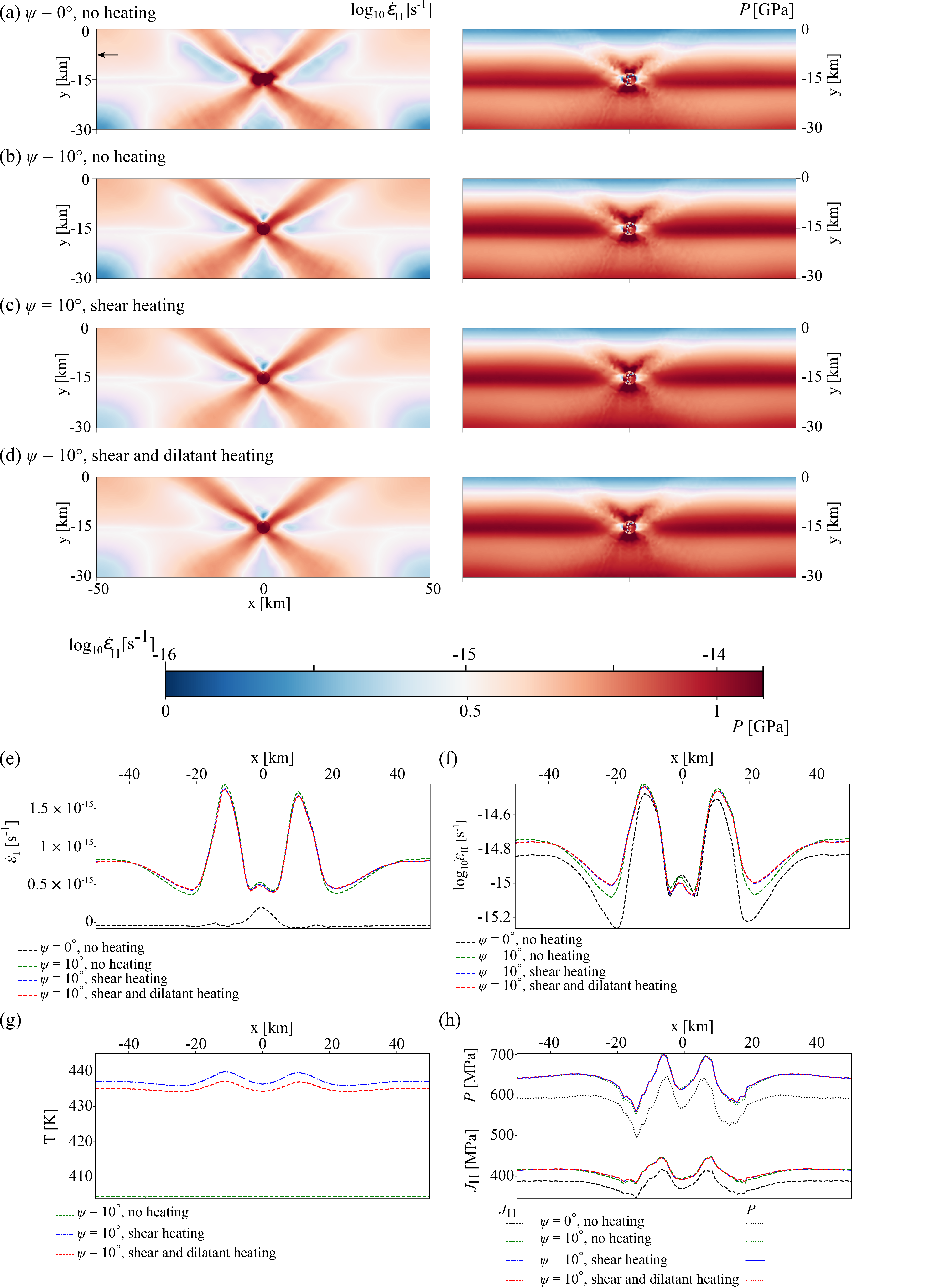}
\caption{TEST 2 results shown in undeformed configuration from the benchmark setup of 100 km ${\times}$ 30 km (Figure \ref{duretz20setup}). The left panel from (a) to (d) shows the logarithm of the second invariant of strain rate; while the right panel indicates the pressure. Model with: (a) zero dilatancy without dissipative heating, (b) 10${^\circ}$ dilatancy without dissipative heating, (c) 10${^\circ}$ dilatancy with shear heating, (d) 10${^\circ}$ dilatancy with shear and dilatant heating, The arrow at 7.5 km below the surface of the model in (a) indicates the starting location where horizontal profiles were extracted: (e) volumetric strain rate, (f) logarithm of deviatoric strain rate, i.e., second invariant of strain rate, (g) temperature without heating, with shear heating alone and shear plus dilatant heating, and (h) pressure and the second invariant of deviatoric stress.}
\label{duretz20results}
\end{figure}
{{\begin{table*}
\begin{minipage}{160mm}
\caption{TEST 2: rheological parameters used in the benchmark setup shown in Figure \ref{duretz20setup}. Bulk modulus, ${K=5\times 10^{10}}$ Pa, ${C_p}$ = 1050 J kg${^\textrm{-1}}$K${^\textrm{-1}}$, ${\rho}$ = 2700 kgm${^\textrm{-3}}$ and a background strain rate was set at ${10}$\textsuperscript{-15} s\textsuperscript{-1}. The dilatancy angle ${\psi\;(^\circ)}$ ranges between 0 and 10${^\circ}$ for the matrix and was kept at 0${^\circ}$ for the inclusion (see text for details). These parameters were drawn from Duretz et al. \protect\shortcite{duretz2020}.}
\label{tabb}
\begin{center}
\begin{tabular}{@{}llccccccl}
\hline
Material & ${A}$ (Pa${^{-m}}$s${^\textrm{-1})}$ & ${m}$ & $E_a\; $ (kJ mol${^\textrm{-1})}$& ${\alpha_{th}}$ (m${^\textrm{2}}$s${^\textrm{-1}}$)&${\phi\;(^\circ)}$ & ${c_0}$ (MPa) \\
\hline
Matrix&3.16${\times}$10${^\textrm{-26}}$&3.3&186.5&${8.82\times{10^{-7}}}$ &30  & 50\\
Inclusion&1${\times}$10${^\textrm{-20}}$&1&0&${8.82\times{10^{-7}}}$ &0 &0.1\\
\hline
\end{tabular}
\label{duretz2020table}
\end{center}
\end{minipage}
\end{table*}}}

\subsubsection{Strain localization in brittle and ductile regimes}

The models were run for 1.2811 Myrs (${4.04\times10^{13}}$ s) with an adaptive time stepping. The initial boundary strain rate and the time of the experiment assure a shortening of only 4\%, consistent with our assumption of small strain theory. We carried out simulations with zero dilatancies without heating (Figure \ref{duretz20results}a), 10${^\circ}$ dilatancy without heating (Figure \ref{duretz20results}b), with shear heating (Figure \ref{duretz20results}c) and with shear and dilatant heating (Figure \ref{duretz20results}d).  Whether viscoelastic or viscoplastic deformation dominated at any time step, both regimes were characterized by the formation of shear bands initializing within the weak zone and propagating into the brittle and the ductile domains. 

The upper brittle domain conformed to a viscoplastic deformation, characterized by narrow shear bands in all simulation cases; while the lower ductile domain conformed to a viscous deformation, characterized by comparatively broader shear bands which propagated towards the bottom boundaries (Figure \ref{duretz20results}a-d). While the shear bands attenuated in both upper and lower domains towards the boundaries, the attenuation occurred faster in the hotter ductile regions. 
The transition from brittle to ductile behaviour was observed within a depth range of 14.21 km and 17.7 km, characterized by a distinct horizon and peak pressure, except within the initial weak zone where pressure remains minimal (Figure \ref{duretz20results}a-d).

Our result with ${\psi=}$ 10${^\circ}$ succinctly captured similar features to the work of Duretz et al \shortcite{duretz2021} with ${10^\circ}$ dilatancy angle and bulk modulus of ${5 \times 10^{10}}$ Pa, despite using different numerical discretization schemes and constitutive law updates. These features include: (1) the vertical symmetry of the deformation in terms of the propagation and position of shear bands (upper versus lower domains), (2) the style of dominant shear bands (narrower in the brittle regime and broader in the ductile regime), (3) ${\sim}{33^\circ}$ angle of shear bands compared to the range of ${\sim}{30^\circ}$ and ${35^\circ}$ of the reference model \cite{duretz2021}, corresponding to theoretical Arthur or intermediate mechanically stable shear band angles defined by ${45^\circ\pm(\varphi+\psi)/2}$ \cite{arthur1977,kaus2010}, and (4) the brittle-ductile transition (just below the weak inclusion), amplitudes of deformation as well as the pressure evolution (Figure \ref{duretz20results}c). 

Including heat production (either shear heating or shear and dilatant heating were accounted for) smoothes out the shear bands in the brittle and ductile domains (Figures \ref{duretz20results}c and \ref{duretz20results}d). In terms of 1-D comparisons of volumetric strain rate (calculated as ${\dot{\varepsilon}_{kk}}$) and deviatoric strain rate invariant, we observe that they are not affected by either shear heating or shear and dilatant heating (Figure \ref{duretz20results}e-f), except that dilatant strain rates are higher for non-zero dilatancy and marginally higher for 10${^\circ}$ dilatancy without heating. Localization is higher for non-zero dilatancy (Figure \ref{duretz20results}f). We also observed a temperature increase of between 31.42 K and 35.32 K for shear heating and 29.73 K and 32.66 K for shear and dilatant heating corresponding to a temperature reduction of between 1.69 K and 2.66 K. The difference between the second invariant of stress between zero-dilatancy and 10${^\circ}$ dilatancy is between 16.38 MPa and 39 MPa, while the pressure difference ranged between 30 MPa and 75.2 MPa (Figure \ref{duretz20results}e-f). Despite the constant background strain rate in these simulations, dilatant plastic feedback through dilatancy has a small effect on the temperature evolution. 

\subsection{The role of plastic dilatancy}
\begin{figure*}
    \centering
\includegraphics[width=0.75\textwidth]
{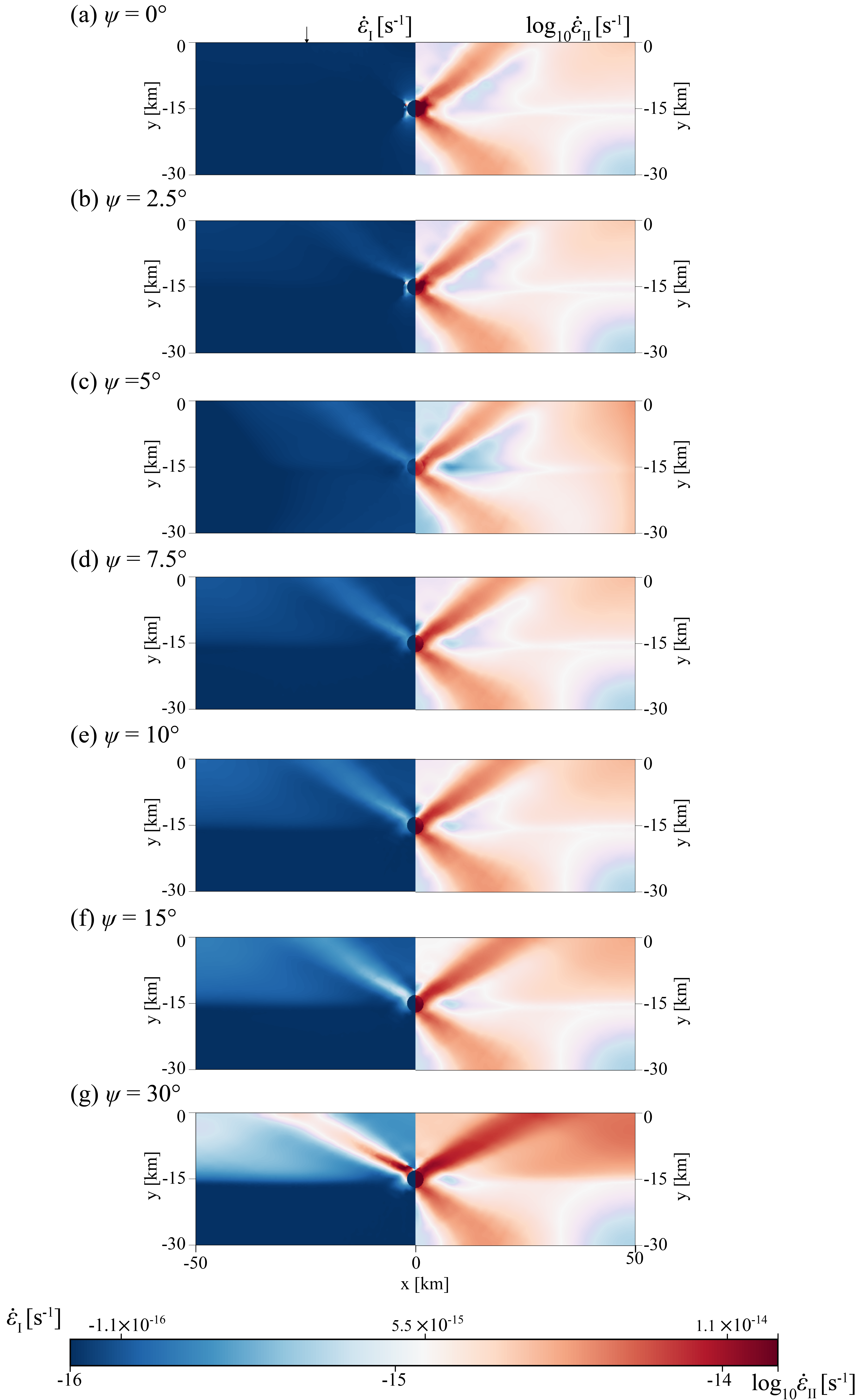}
\caption{The effect of constant plastic dilatancies on volumetric strain rate invariant (left half panel from x = 0) and deviatoric strain rate invariant (right half panel from x = 0) where shear and dilatant plastic deformations contribute to dissipative heating. The dilatancy angles span (a) 0${^\circ}$ (non-dilatant plasticity) to (g) 30${^\circ}$ (an extreme case of the friction angle = dilatancy angle). The black arrow indicates the location where 1-D profiles were extracted and shown in Figure \ref{duretz2021resultsdilatancy1D}. We have shown one half of each quantity due to the vertical symmetry about x = 0.}
\label{duretz2021resultsdilatancy}
\end{figure*}

\begin{figure*}
    \centering
\includegraphics[width=0.75\textwidth]
{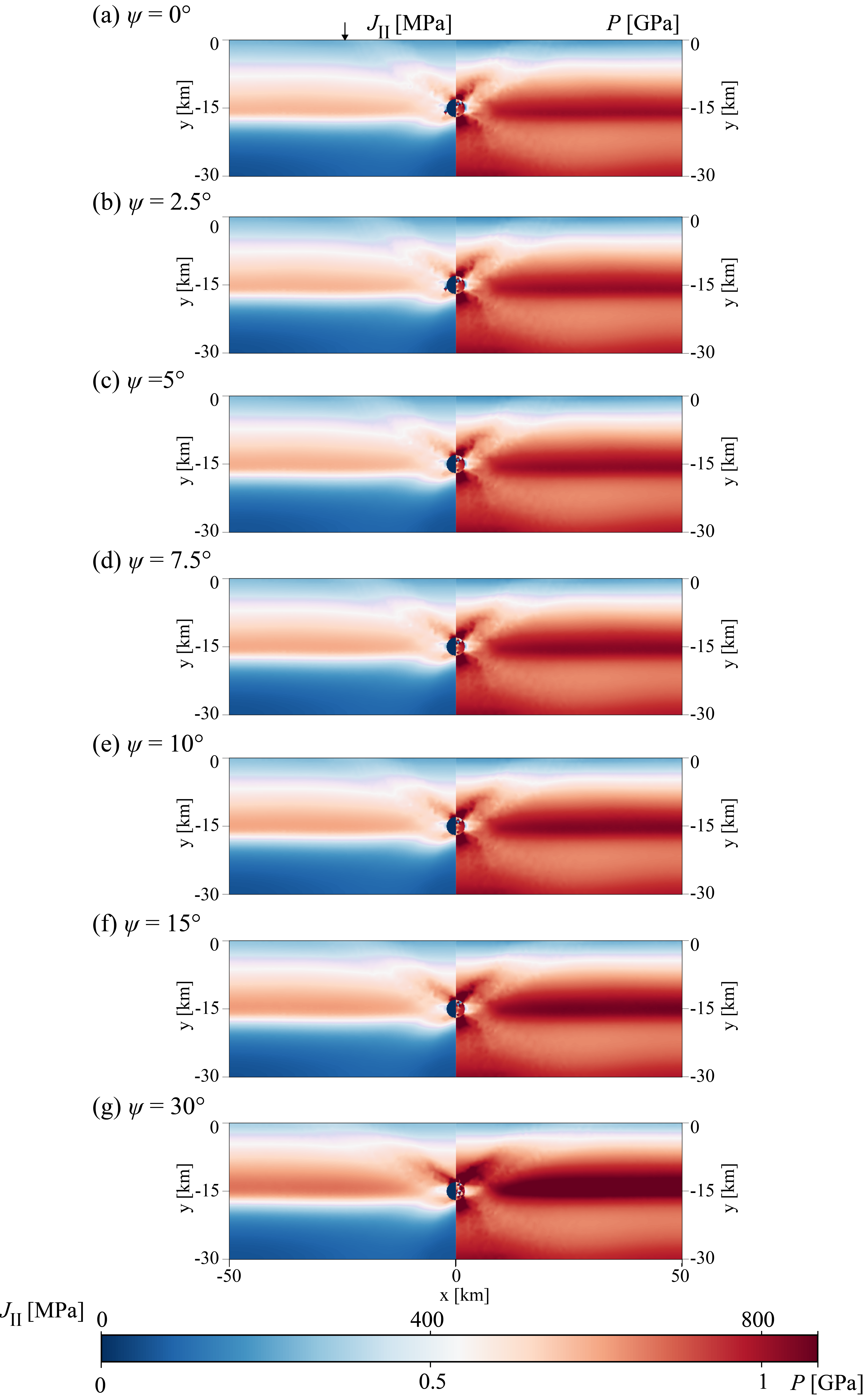}
\caption{The effect of constant plastic dilatancies on deviatoric stress invariant (left panel) and pressure (right panel) where shear and volumetric plastic deformations contribute to dissipative heating. The dilatancy angles span (a) 0${^\circ}$ (non-dilatant plasticity) to (g) 30${^\circ}$ is (an extreme case of the friction angle = dilatancy angle). The black arrow in (a) indicates the location where 1-D profiles were extracted and shown in Figure \ref{duretz2021resultsdilatancy1D}.}
\label{duretz2021resultsdilatancyJIIP}
\end{figure*}

Since the contribution to volumetric stresses and strains includes dilatant plasticity, we investigated the effect of varying the dilatancy angles, for a given value of elastic compressibility (${K\;=\;5\;\times\;10^{10}}$ Pa). We set the  activation energy to zero for the weak inclusion to ensure viscous behaviour therein irrespective of temperature (Table \ref{tabb}). The results for constant dilatancy angles ${\psi = 0^\circ\;\mathrm{to}\;30^\circ}$ are shown in Figure \ref{duretz2021resultsdilatancy} for volumetric and deviatoric strain rate invariants, and Figure \ref{duretz2021resultsdilatancyJIIP} for deviatoric stress invariant and pressure.

While the volumetric strain rate invariant increased in amplitude from non-dilatant plasticity to the extreme case of associated plasticity with dilatancy angle = friction angle = $30{^\circ}$, we observed uniform volumetric strains in the ductile domain (below 15 km from the surface) as shown in Figure \ref{duretz2021resultsdilatancy}. {Concerning the} the deviatoric strain rate invariant, the shear bands are narrow in the brittle domain (above 15 km) and broader in the ductile domain (below 15 km), as shown in Figure \ref{duretz2021resultsdilatancy}. Increasing dilatancy angles leads to increasing strain magnitudes within the shear bands in the brittle domain.

The evolution of the deviatoric stress invariant and the pressure indicate a gradual increase in both quantities near the brittle-ductile transition as the dilatancy angles increased, with the pressure saturating around the brittle-ductile transition for dilatancy of 30${^\circ}$ (Figure \ref{duretz2021resultsdilatancyJIIP}). The zone of highest deviatoric stress invariant and pressure also moved upwards as dilatancy angles increased.

\begin{figure*}
    \centering
\includegraphics[width=12cm]
{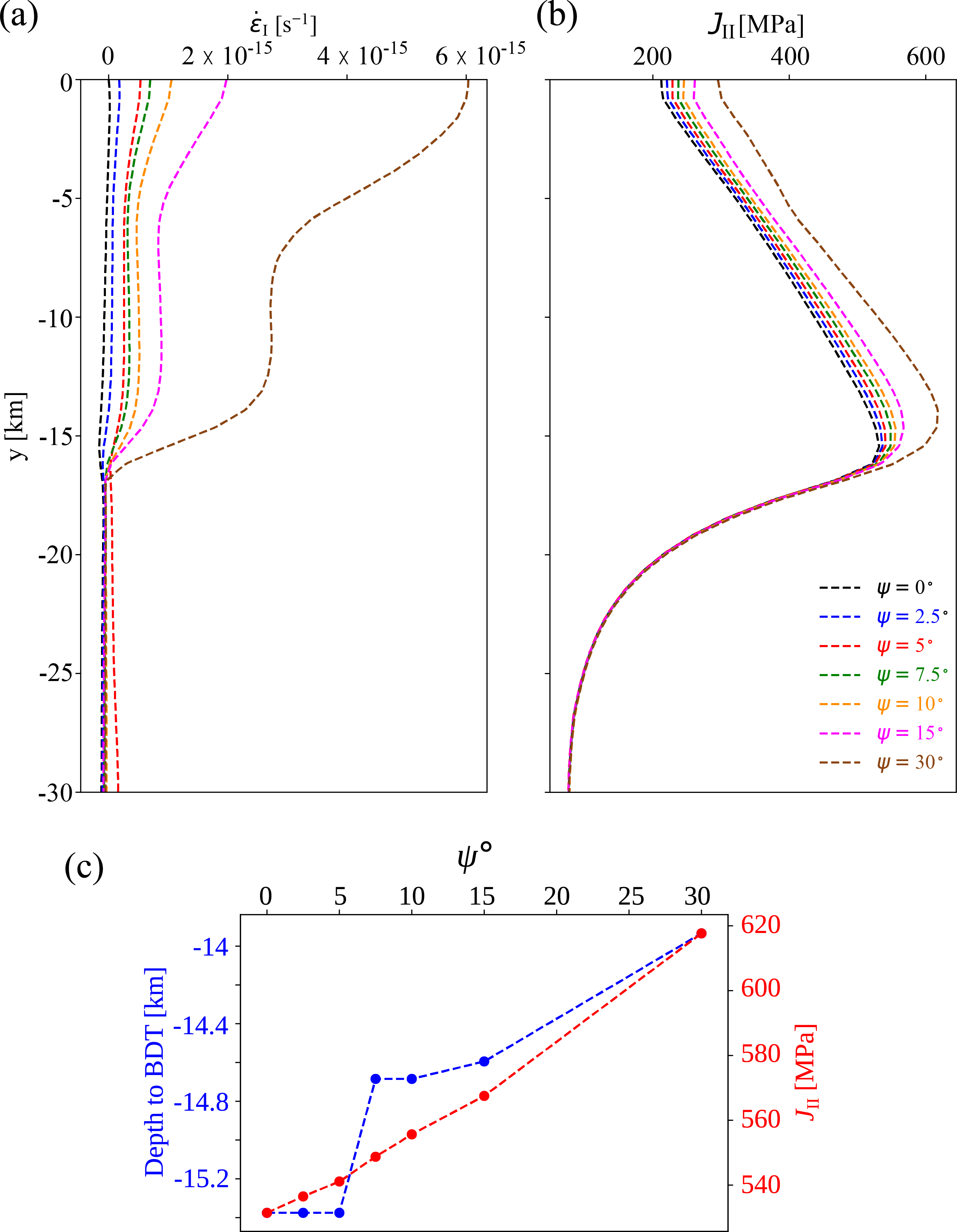}
 \caption{1-D plots of (a) volumetric strain rate invariant and (b) deviatoric stress invariant with locations indicated by arrows in Figure \ref{duretz2021resultsdilatancy}a and Figure \ref{duretz2021resultsdilatancyJIIP}a, (c) plot of the maximum deviatoric stress invariant (used as the limit of the brittle domain) and depth to the brittle-ductile transition with respect to the dilatancy angles.}
\label{duretz2021resultsdilatancy1D}
\end{figure*}

To further assess the evolution of deformation and stress state, we compared the evolution of the volumetric strain rate invariant and deviatoric stress invariant, we extracted 1-D curves for both quantities (Figure \ref{duretz2021resultsdilatancy1D}). Due to the vertical symmetry of deformation,
we displayed plots at the location indicated in Figure \ref{duretz2021resultsdilatancy}a and Figure \ref{duretz2021resultsdilatancyJIIP}b. Concerning the volumetric strain rate invariant, we observed that increasing dilatancy angles leads to an increase in the volumetric strain rate invariant in the brittle domain (Figure \ref{duretz2021resultsdilatancy1D}a). With respect to the second invariant of the stress tensor (${J_\mathrm{II}}$), the curves intersect near the brittle-ductile transition below 15 km depth (Figure \ref{duretz2021resultsdilatancy1D}b). A first-order observation is the similarity in the shapes of the curves. The maximum deviatoric stress invariant ranged from 531 MPa for ${\psi = \mathrm{0}^\circ}$ and 618 MPa for ${\psi = \mathrm{30}^\circ}$ at depths of 15.4 km and 13.9 km, respectively (Figure \ref{duretz2021resultsdilatancy1D}c). This indicates that higher dilatancy angles require higher stress at a relatively shallower depth. Even though the experiment is a pure shear case, the impact of dilatancy is more obvious in the brittle domain. Decreasing dilatancy angles from realistic values of 10${^\circ}$, i.e., at least 20${^\circ}$ less than the friction angle \cite{vermeer1984} to non-dilatant plasticity led to a 25 MPa (or 10${\%}$) reduction in the deviatoric stress invariant near the surface, and ${\sim}$30 MPa (or 5${\%}$) reduction at ${\sim}$14.5 km before the brittle-ductile transition.

If we use this second stress invariant as a proxy for the magnitude of shearing exerted by a state of stress in the medium\ \cite{bower}, we may argue that lower dilatancy angles reduce the magnitude of shear stresses in the medium compared to higher dilatancy angles. The synchrony of the deviatoric stress invariants for different dilatancy angles in the ductile domain can be attributed to the assumption that there are no volumetric plastic deformations in the ductile domain, per our constitutive description. The apparent attenuation in the second stress invariant for small dilatancy angles may explain an increase in strain rate, alluding to the inverse relationship between the strain rate and second stress invariant (Equation \ref{vpupdate}).

\section{The effect of volumetric plastic (dilatant) heating}
The pure shear experiment was implemented with the deliberate choice of constant strain rate boundary conditions by adapting the boundary velocity at each time step, and reproducing the conditions for benchmarking purposes \cite{duretz2021}. We showed that the contribution of dilatant plastic dissipation to volumetric or deviatoric localization was not significant (Figure \ref{duretz20results}e-f) and impacted the thermal state by reducing the temperature by up to 3 K (Figure \ref{duretz20results}g). 

To investigate how different boundary conditions may influence deformation and heat production when dilatant plastic dissipation is included, we ran another series of simulations: where we imposed constant boundary velocities (TEST 3), utilized an isothermal domain (TEST 4A), and  high-strain rate experiments (TEST 4B). 

\subsection{TEST 3: Nucleation zone in brittle-ductile domain}

\begin{figure*}
    \centering
\includegraphics[width=16cm]{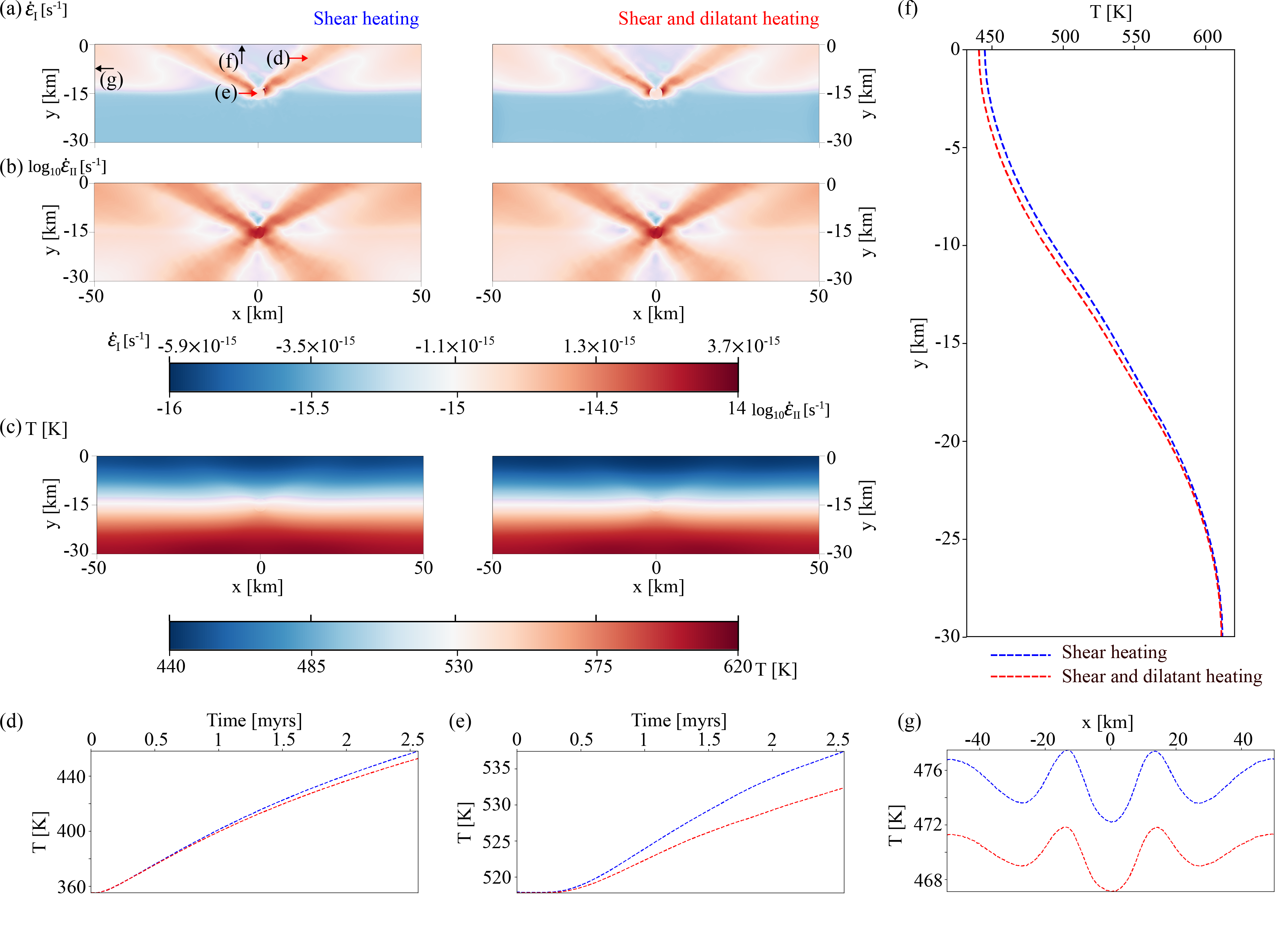}
\caption{TEST 3: Results of deformation with dissipative feedback from shear heating and shear with dilatant heating. (a) volumetric strain rate invariant (b) logarithm of the deviatoric strain rate invariant, (c) evolved temperature at the end of the experiment, (d) temperature evolution with time for an element within the shear band, (e) temperature evolution with time for an element within the weak inclusion, (f) vertical temperature profile near the weak inclusion crossing one of the shear band, and (g) horizontal temperature profile in the brittle domain along the model. The red arrowheads in (a) are locations where the temperature histories were tracked within the shear band and weak inclusion, respectively; while the black arrowheads indicate starting locations where 1-D temperature profiles were extracted. The left panel in Figure \ref{duretz2021breakingthesymmetry2D}a-c represents results from shear heating, while the corresponding right panels show the results when dilatant heating is included.}
\label{duretz2021breakingthesymmetry2D}
\end{figure*}

In this test case 3, we utilized the same input setup shown in Figure \ref{duretz20setup} with the same material parameters. In addition, we replace the varying velocity boundary conditions with constant boundary velocities. We ran the simulations for 2.5622 myrs (${8.08\times10^{13}}$ s) for a target shortening of 8\%. The results with shear heating alone, and shear with dilatant plastic feedback are shown in Figure \ref{duretz2021breakingthesymmetry2D}.

Shown in Figures \ref{duretz2021breakingthesymmetry2D}a and \ref{duretz2021breakingthesymmetry2D}b are the volumetric strain rate invariant and deviatoric strain rate invariant, respectively, for cases of shear heating alone and shear with dilatant heating. The corresponding temperatures at the end of the simulations are shown in Figure \ref{duretz2021breakingthesymmetry2D}c. We again note that the deformation is symmetric (vertical symmetry about the initial weak zone depth) similar to the pure shear experiment reported in test TEST 2, with similar temperature evolution in 2-D. While the volumetric strain rate invariant followed the shear bands in the brittle domain, we note the uniformity of the volumetric strain rate invariant in the ductile domain (Figure \ref{duretz2021breakingthesymmetry2D}a), which is consistent with our assumptions of treating volumetric plastic deformations only from brittle plasticity and not ductile creep. The observation of shear bands in the brittle and ductile regimes for the deviatoric strain rate invariant is also consistent with our accounting for deviatoric strains in both the brittle and ductile domains. To compare the temperature, we extracted the temperature evolution for an element within a shear band and within the weak zone (Figures \ref{duretz2021breakingthesymmetry2D}d and \ref{duretz2021breakingthesymmetry2D}e). Within the shear band element, the temperature increased by 102 K for shear heating. We observed that both the shear heating and shear with dilatant heating temperatures evolved similarly until ${\sim}$0.4 million years where the temperatures were offset until the end of the experiment with the inclusion of dilatant heating leading to reduction of 5 K at the end of the simulation. In the case of the element within the weak inclusion, the overall temperature rise was 20 K at the end of the experiment when shear heating was considered alone; however, this temperature rise was reduced by 5 K when dilatant heating was included. The vertical temperature profile also shows that dilatant heating reduced the amount of heat dissipated by shear heating by 4 K (Figure \ref{duretz2021breakingthesymmetry2D}f). The same observation is true for the horizontal profile where the inclusion of dilatant heating reduced the heat produced by shear heating by 5.5 to 6 K (Figure \ref{duretz2021breakingthesymmetry2D}g).
The temperature due to shear dissipative heating versus temperature due to shear and dilatant dissipative feedback indicates that dilatant contributions introduce a less heating effect in the brittle domain.  

\subsection{TEST 4A: Viscoplasticity in an isothermal domain and possible rheological change due to thermal dissipation}\label{TEST4A}
Motivated by the impact of dilatant plastic deformation in thermal dissipation, we returned to the initial setup for viscoelastic benchmarking shown in Figure \ref{duretz14setup}. We utilized the material properties discussed for TEST 3, and we have included dilatancy angle of 10${^ \circ}$. The initial temperature we have used was set at 473 K, to ensure that the impact of ductile creep was suppressed in order to ensure a viscoplastic initial rheology. We have run the simulations for ${4\times10^{12}}$ s (126 kyrs).

\begin{figure}
    \centering
\includegraphics[width=8cm]{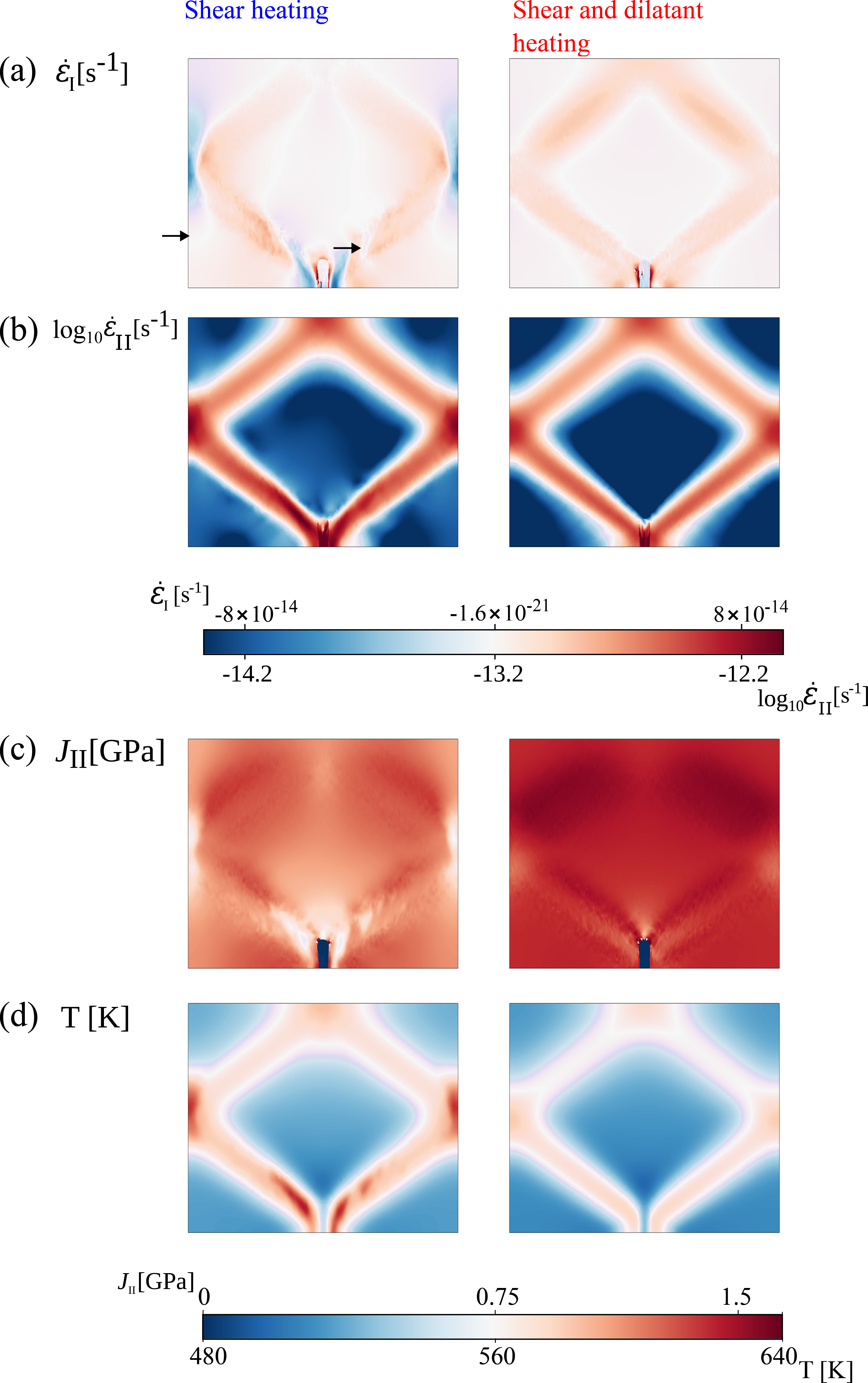}
\caption{TEST 4A results shown in deformed configuration for a setup shown in Figure \ref{duretz14setup}. We have utilized an initial isothermal temperature of 473 K to ensure viscoplastic deformations. The results show two columns indicating the influence of shear heating, and shear with dilatant heating, respectively: (a) volumetric strain rate invariant (b) deviatoric strain rate invariant, (c) second invariant of stress, and (d) temperature evolution at the end of the simulation. These results are plotted in deformed configuration measuring 57.4 km in the \textit{x}-direction and 48.8 km in the \textit{y}-direction similar to the same observation in Figure \ref{duretz14results1}. The arrows on the left border and one element within a shear band in (a) are pointers to, locations where 1-D profiles and some history-dependent variables were extracted and shown in Figure \ref{horizontalprofilesacrossshearbands}a-e, and Figure \ref{horizontalprofilesacrossshearbands}f-j, respectively.}
\label{vevpduretz2014}
\end{figure}

We observed differences in the volumetric strain rate invariant (Figure \ref{vevpduretz2014}a), deviatoric strain rate invariant (Figure \ref{vevpduretz2014}b), the deviatoric stress invariant (Figure \ref{vevpduretz2014}c) and the temperature at the end of the simulations (Figure \ref{vevpduretz2014}d). The shear bands were observed at an angle of ${\sim}$35-42${^\circ}$. Utilizing a friction angle of 30${^\circ}$ and dilatancy of 10${^\circ}$, the angle of the observed shear bands within the matrix corresponds to Arthur or intermediate angles, suggesting that plasticity parameters influenced the behaviour of shear bands during viscoplastic flow as opposed to the ductile shear bands which are not influenced by these parameters.

When dilatant heating was included, shear bands were smoother as seen in the volumetric strain rate invariant and deviatoric strain rate invariant (Figures \ref{vevpduretz2014}a and \ref{vevpduretz2014}b). While the lowest deviatoric stress invariant was observed in the weak inclusion whether shear heating or shear and dilatant heating were considered, we observed a higher deviatoric stress invariant when dilatant heating was included (Figure \ref{vevpduretz2014}c). The temperature at the end of the simulation showed that temperatures increased for both shear heating and shear with dilatant heating (Figure \ref{vevpduretz2014}d). However, temperatures were higher within the shear bands when shear heating was considered alone compared to when dilatant heating was included (Figure \ref{vevpduretz2014}d). Our deduction from these results is that including the dilatant term contributed to a higher deviatoric stress invariant, and lower deviatoric strain rate invariant with less heating effect within the shear bands.

\begin{figure}
    \centering
\includegraphics[width=15cm]{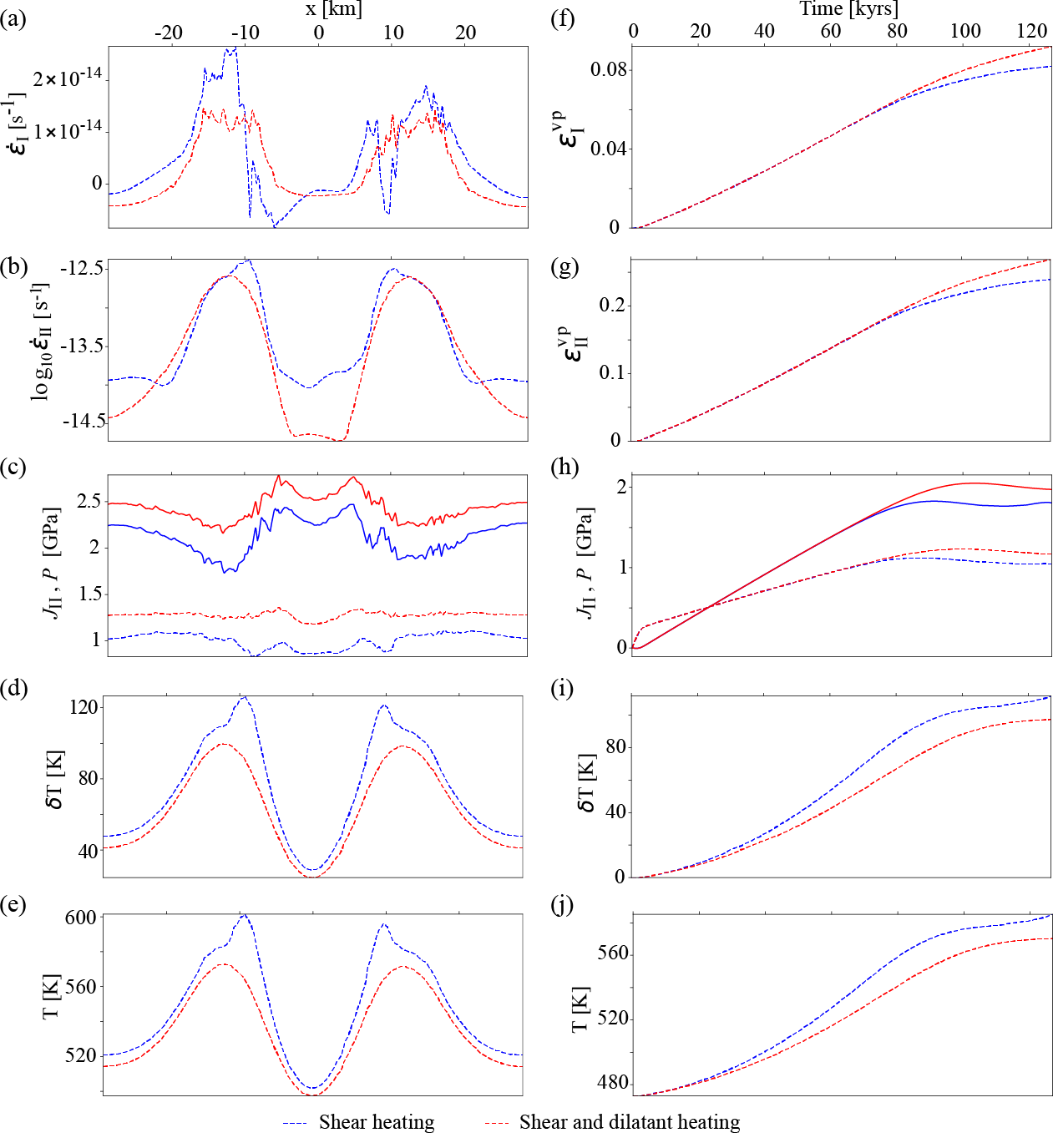}
\caption{TEST 4A (a-e) are 1-D profiles extracted across two dominant shear bands from location indicated by arrow on the left border of the model shown in Figure \ref{vevpduretz2014}a: showing (a) volumetric strain rate invariant (b) deviatoric strain rate invariant, (c) second invariant of stress shown in dashed lines, and pressure shown in full lines, (d) temperature difference between the final temperature and initial temperature, and (e) final temperature; (f-j) are 1-D history evolution of quantities extracted from an element within one of the dominant shear bands whose location is shown in Figure \ref{vevpduretz2014}a: (f) volumetric viscoplastic strain invariant, (g) deviatoric viscoplastic strain invariant, (h) deviatoric stress invariant shown in dashed lines and pressure shown in full lines, (i) evolution of temperature, and (j) integrated temperature.}
\label{horizontalprofilesacrossshearbands}
\end{figure}

To further illustrate how dilatant plastic feedback may affect the behaviour of the plotted variables in Figure \ref{vevpduretz2014}, we extracted 1-D plots and history plots from locations shown in Figure \ref{vevpduretz2014}a. Across the model, the contribution of dilatant plastic heating is either a reduction or an increase in the volumetric strain rate invariant (Figure \ref{horizontalprofilesacrossshearbands}a) and the deviatoric strain rate invariant (Figure \ref{horizontalprofilesacrossshearbands}b). We further confirmed that including dilatant heating led to increased pressure and deviatoric stress invariant in the brittle domain, with the pressure reducing within the shear bands for both shear heating and shear  with dilatant heating between x = -7.5 km and -20 km (Figure \ref{horizontalprofilesacrossshearbands}c). This observation is consistent with our correction for pressure during viscoplastic flow (Equation \ref{pressure_update}). That is, zones with the highest strains (volumetric or deviatoric) follow from zones with the highest viscoplastic multiplier (${\Delta\gamma^\textrm{vp}}$), hence the more pressure correction. The pressure difference ranges between 202 MPa and 509 MPa when dilatant heating is included, while the difference in deviatoric stress invariant ranges is between 187.82 MPa and 435 MPa. Across the profile, the temperature rise was between 28.9 K and 126.6 K when shear heating was considered alone and between 24.5 K and 99.6 K when dilatant heating was included with shear heating (Figure \ref{horizontalprofilesacrossshearbands}d). The temperature reduction when dilatant heating was included was between 4.3 K and 38 K (Figure \ref{horizontalprofilesacrossshearbands}e). 

In Figure \ref{horizontalprofilesacrossshearbands}f-j, we show the history-dependent evolution of some of the variables for an element within the shear band whose location was shown in Figure \ref{vevpduretz2014}a, near the weak zone. We observed that the integrated dilatant and deviatoric viscoplastic strain invariants indicate that dilatant heating combined with shear heating influences both quantities (Figures \ref{horizontalprofilesacrossshearbands}f and \ref{horizontalprofilesacrossshearbands}g). This indicates that at longer simulation times, dilatant heating may increase irreversible strains. This is consistent with the along-axis profile where it was observed that when dilatant heating was included, it led to an increase or decrease of volumetric strain rate or deviatoric strain rate as shown in Figure \ref{horizontalprofilesacrossshearbands}a-b. The evolution of pressure and deviatoric stress invariant approached a steady state behaviour at ${\sim}$83.6 kyrs (Figure \ref{horizontalprofilesacrossshearbands}h) at which point the inclusion of dilatant heating offset the temperature by about 18 K (Figures \ref{horizontalprofilesacrossshearbands}i-j). While the irreversible strains, pressure and stress invariants were similar until ${\sim}$76 kyrs, the temperature for either: shear heating alone and shear with dilatant heating were being offset from 20 kyrs and the offset increased as the simulations proceeded. 

It can be argued that shear heating alone can enhance deviatoric strain localization as already shown in previous studies \cite{thielmann2012}, while shear with dilatant heating also enhances deviatoric strain localization even if delayed (Figure \ref{vevpduretz2014}b), increases deviatoric stress invariant (Figure \ref{vevpduretz2014}c) consistent with an inverse relationship between strain rate and deviatoric stress invariant (Equation \ref{vpupdate}) and reduce the amount of heat produced (Figure \ref{vevpduretz2014}d). Note that the same boundary conditions used here ensure a constant strain rate, which suggests that even for nominally volume-preserving boundary conditions, volumetric plastic effects contribute to the rate of deformation and temperature evolution depending on the rheology. It must also be noted that as temperature increases due to heating feedback, the likelihood of the rheology switching from viscoplastic to viscoelastic is not ruled out. In fact, in some parts of the model, this rheological switch was already seen when the material stopped yielding, i.e., in these areas where the Drucker Prager criterion (Equation \ref{drucker}) was no longer satisfied.

\subsection{TEST 4B: Viscoelasticity and viscoplasticity at high strain rates}\label{TEST4B}
\begin{figure}
    \centering
\includegraphics[width=14cm]{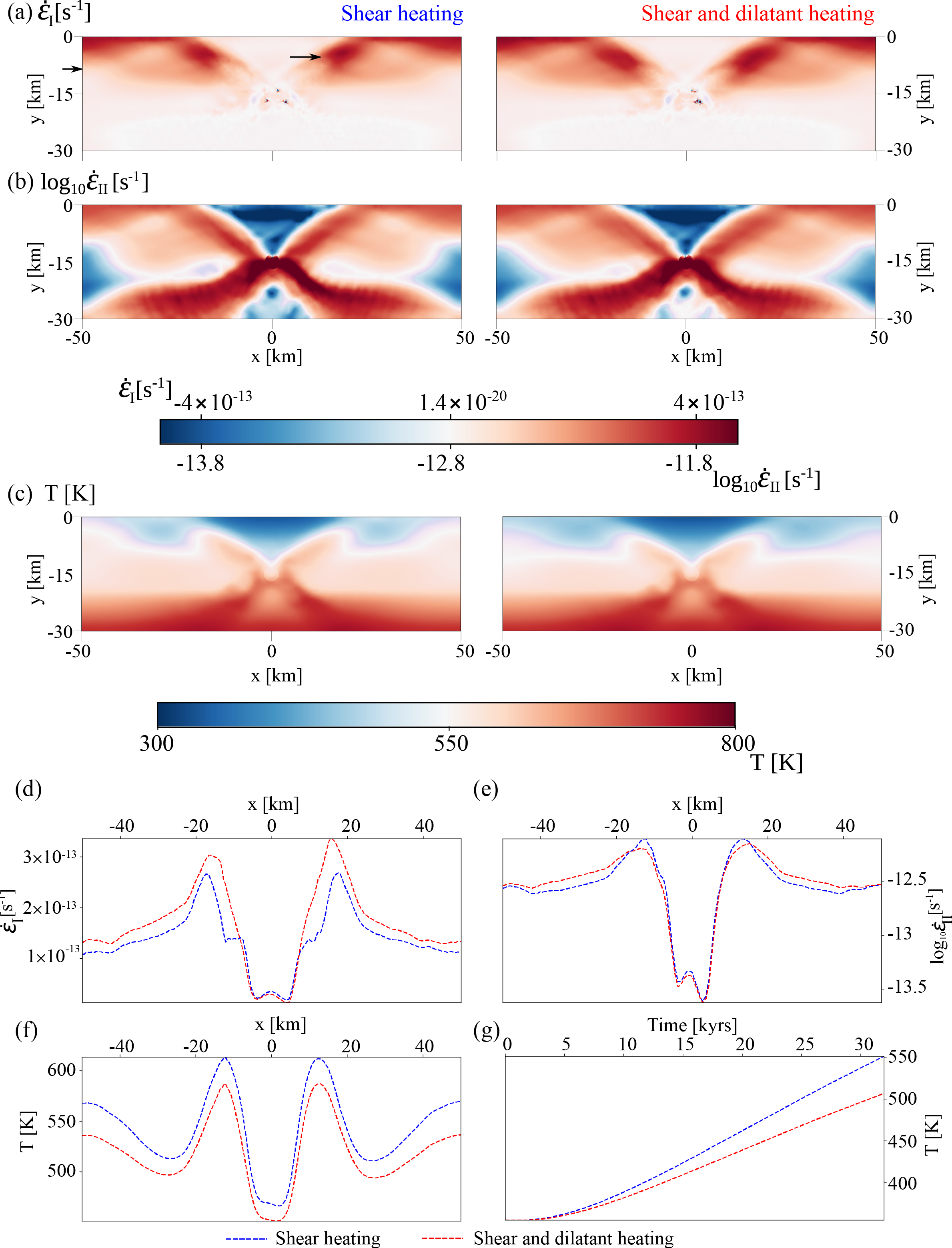}
\caption{TEST 4B comparison of deformation states when accounting for shear heating alone and when dilatant plastic contributions were included (see Section \ref{TEST4B}). (a) volumetric strain rate invariant, (b) deviatoric strain rate invariant, and (c) accumulated temperature. 1-D profiles extracted from ${\sim}$7.5 km (indicated by an arrow on the border of the model shown in Figure \ref{highstrainrate}a) for: (d) volumetric strain rate invariant, (e) deviatoric strain rate invariant, and (f) corresponding temperature across the profile. (g) temperature history for an element within the shear band in the brittle part of the model domain (location of the element tracking the temperature within a shear band is indicated by an arrow in Figure \ref{highstrainrate}a).}
\label{highstrainrate}
\end{figure}

To perform experiments at high strain rates, we returned to the setup shown in Figure \ref{duretz20setup} where we increased the initial boundary strain rate by increasing the initial velocity by three orders of magnitude to ${1\times10^{-8}\;\mathrm{m/s}}$ for the vertical boundaries and to ${3\times10^{-9}\;\mathrm{m/s}}$ for the bottom boundary compared to TEST 3. The total amount of shortening was 24\%. We ran the simulations for ${1\times10^{12}}$ s (${\sim}$32 kyrs) and show the results in Figure \ref{highstrainrate}a-c and associated 1-D profiles for temperature variation across the brittle domain and temperature history within an element in the shear band are shown in Figure \ref{highstrainrate}d-g. 
The shear bands initiated within the weak zone and propagated through the brittle and ductile domain, secondary shear bands appeared on the top boundary as shown in Figures \ref{highstrainrate}a and \ref{highstrainrate}b. We also observed that the ductile shear bands had higher amplitudes in the deviatoric strain rate invariant than brittle shear bands (Figures \ref{highstrainrate}b). We also followed the deformed zones where the heat is produced as shown in the temperature evolution (Figure \ref{highstrainrate}c) where more heat was produced in the ductile domain than the brittle domain. 

In assessing the discrepancy in the deformation and the temperature when shear heating is considered alone or dilatant plastic contribution is incorporated, we observed from the horizontal profiles crossing two shear bands in the brittle domain that the feedback from dilatant and shear heating led to increased volumetric strain rate (Figure \ref{highstrainrate}d) and a slight increase or decrease in the deviatoric strain rate (Figure \ref{highstrainrate}e). Concerning temperature, the difference is up to 6${\%}$ (Figure \ref{highstrainrate}f), while an example of temperature evolution in an element within the shear band in the brittle domain shows that the dilatant plastic feedback can lead to an 8${\%}$ reduction in temperature after 30,000 years. These results demonstrate, as with previous test cases, the ability of dilatant plastic dissipation to potentially slow down shear heating and impacts on the volumetric strain rate. Over long time scales, this effect is expected to be more dominant. 

{\section{Application to Lithospheric Dynamics}

To further illustrate the potential use of our proposed viscoelastic-viscoplastic constitutive law, we carried out simulations that tackle the lithospheric-scale.
\subsection{Initial model set-up and boundary conditions}
We investigated two models, both composed of a 12 km oceanic crust and a mantle: (1) without a weak zone, and (2) with an elliptical weak zone with major and minor radii 20 km and 8.4 km respectively placed at 32 km below the surface. An example of the model set up with an elliptical weak zone is shown in Figure \ref{input_geo}. The model setup without the weak zone  was aimed at investigating possible lithospheric buckling (or warping) when the model is subjected to long-term compression, while the model with the elliptical weak zone was to preferentially seed localization and to investigate the role of dilatant plastic and shear dissipation compared to shear dissipation alone. The material properties utilized are shown in Table \ref{geodynsubd}. 

The initial vertical thermal distribution in the mantle is computed using the cooling of a semi-infinite half-space model for a 50 myr old lithosphere \cite{turcotte2002,geryabook}:
\begin{equation}
T_y = T_1+(T_0-T_1)\left(1-\dfrac{2}{\pi}\arctan\left[\dfrac{y}{\sqrt{\alpha_\mathrm{th}\tau}}\left(1 + \left(\dfrac{y}{2\sqrt{\alpha_\mathrm{th}\tau}}\right)^4\right)\right]\right),
\end{equation}
where ${T_y}$ is the temperature at a given depth, ${y}$, ${T_0}$ is the temperature at the surface, ${T_1}$ is the temperature at the base of the model; the function in parenthesis approximates the error function, where ${\alpha_\mathrm{th}}$ represents thermal diffusivity, and ${\tau}$ represents the age of the lithosphere in seconds.
The initial temperature at the top of the model is set at 273 K, while the temperature at the base is set at 1673 K:  Both the left and right boundaries have zero heat flux. The top boundary is a free surface, and the bottom boundary is maintained free slip. We apply normal inward constant velocities on both vertical boundaries of the model at a rate of 2 cm/year for at least 5 million years (1.5768 ${\times 10^{14}}$ s). We meshed the uppermost part of the model with 2.5 km element size and increased the element size towards the base of the model.    
\begin{figure*}
    \centering
\includegraphics[width=13cm]{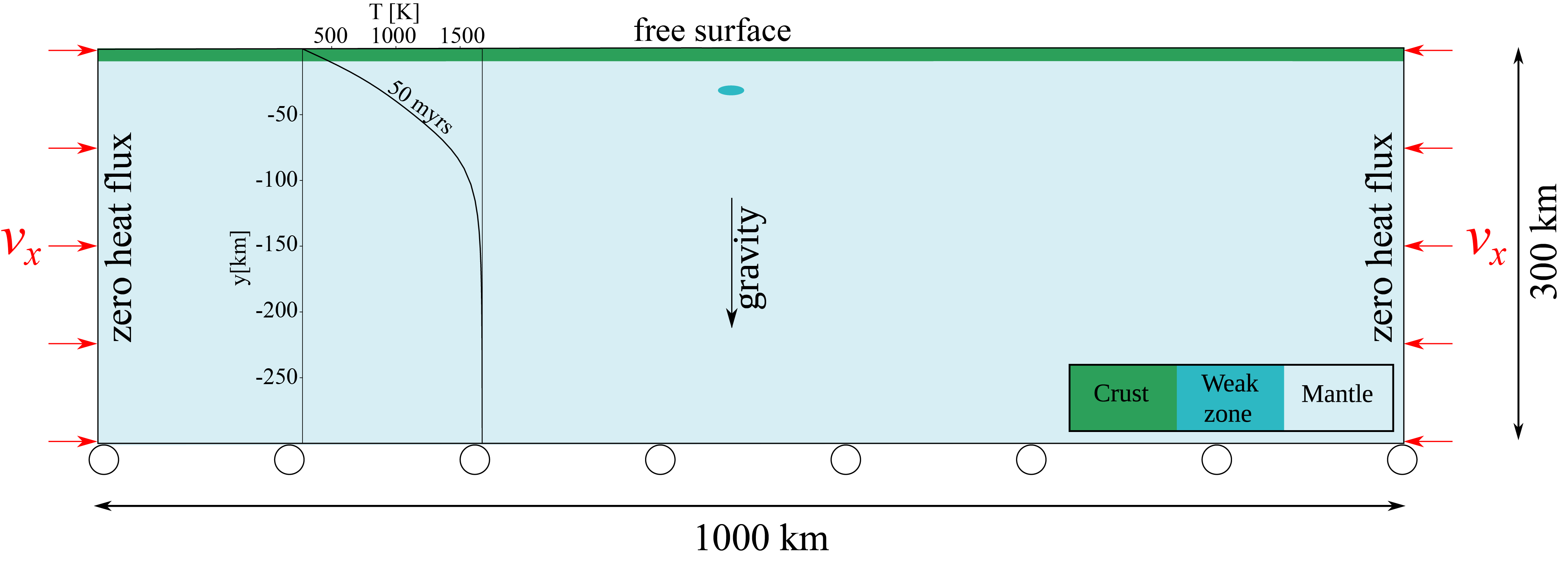}
\caption{Illustrative model setup and boundary conditions for two cases: case 1 without an initial weak zone, and case 2 with an elliptical weak zone. Only case 2 is shown in this sketch. The horizontal compression velocity is 2 cm/year.}
\label{input_geo}
\end{figure*}

{{\begin{table}
\begin{minipage}{150mm}
\caption{Material properties in the geodynamic model. \hspace{2mm}${E}$ (Young's modulus), $v $ (Poisson's ratio), $K$ (bulk modulus), $G$  (shear modulus), $\rho$  (density), $g$ (gravitational constant), ${C_p}$ (specific heat capacity), $\alpha_\textrm{th}$  (thermal diffusivity), ${A}$ (pre-exponential multiplier), $E_a$  (creep activation energy), $R$ (molecular gas constant), ${\varphi}$ (friction angle), ${\psi}$ (dilatancy angle), $c_0$ (initial cohesion), $m$ (stress exponent), and ${1/\mu}$ (relative rate of viscoplastic strain). The references for material parameters were drawn from \textsuperscript{a}Ranalli \protect\shortcite{ranalli}, \textsuperscript{b}Goetze \& Evans \protect\shortcite{goetze79}, \textsuperscript{d}Hirth \& Kohlstedt \protect\shortcite{hirth}
and Thielmann \& Kaus \protect\shortcite{thielmann2012}.}
\label{geodynsubd}
\begin{tabular}{@{}lllllll}
\hline
Property  (unit):  &  Mantle\textsuperscript{a,b,c} & Crust\textsuperscript{a} & Weak zone\textsuperscript{a,b,c}\\
\hline
Mechanical &           &      &  \\[2pt]
\hspace{2mm}${E}$ (GPa) & 25     & 25 & 25\\[2pt]
\hspace{2mm}$v$    & 0.25     & 0.25 & 0.25 \\[2pt]
\hspace{2mm}$G {\;} $ (GPa)           & 10     & 10 & 10 \\[2pt]
\hspace{2mm}$K {\;} $ (GPa)           & 16.7     & 16.7 & 16.7\\[2pt]
\hspace{2mm}$\rho$ (kg.m${^\textrm{-3}}$)    & 3300     & 2950 & 3100\\[2pt]
\hspace{2mm}$ g $ (m.s${^{\textrm{-2}}}$)       & 9.8     & 9.8& 9.8
\\[2pt]
\hspace{2mm}$m $  & 3.5     & 2.3    & 4\\[2pt]
\hline
Thermal &            &      \\[2pt]
\hspace{2mm}${C_p}$  (J kg${^\textrm{-1}}$K${^\textrm{-1}}$)       & 1050& 1050 & 1050\\[2pt]
\hspace{2mm}$\alpha_\textrm{th} {\;}$ (m${^\textrm{2}}$s${^\textrm{-1}}$) &  ${8.7\times{10^{-7}}}$     & ${9.7\times{10^{-7}}}$ & ${9.2\times{10^{-7}}}$\\
\hline
Dislocation creep     &         &  \\[2pt]
\hspace{2mm}${A}$ (Pa${^\textrm{-m}}$ s${^\textrm{-1})}$       & $2\times10^{-21}$   &$1.59\times10^{-14}$&$2.08\times10^{-23}$\\[2pt]
\hspace{2mm}$E_a\; $ (kJ mol${^\textrm{-1})}$    & 535     & 154 & 283 \\[2pt]
\hspace{2mm}$R\;$ (J K${^\textrm{-1}}$mol${^\textrm{{-1}}}$)  & 8.31     & 8.31 & 8.31 \\
\hline
Plasticity &     &        &      &  \\[2pt]
\hspace{2mm}${\varphi {\;} (^\textrm{o})}$&  30     & 30 & 15\\
\hspace{2mm}${\psi {\;} (^\textrm{o})}$      & 10     & 10 & 5\\[2pt]
\hspace{2mm}$c_0 {\;} $(MPa)    & 1         & 1 & 1 \\[2pt]
\hspace{2mm}${1/\mu}$ (s${^ {-1}}$)    & ${10^{-15}}$       & ${10^{-15}}$    & ${10^{-15}}$ \\[2pt]
\hline
\end{tabular}\\
\end{minipage}
\end{table}}}
\begin{figure}
    \centering
\includegraphics[width=15cm]{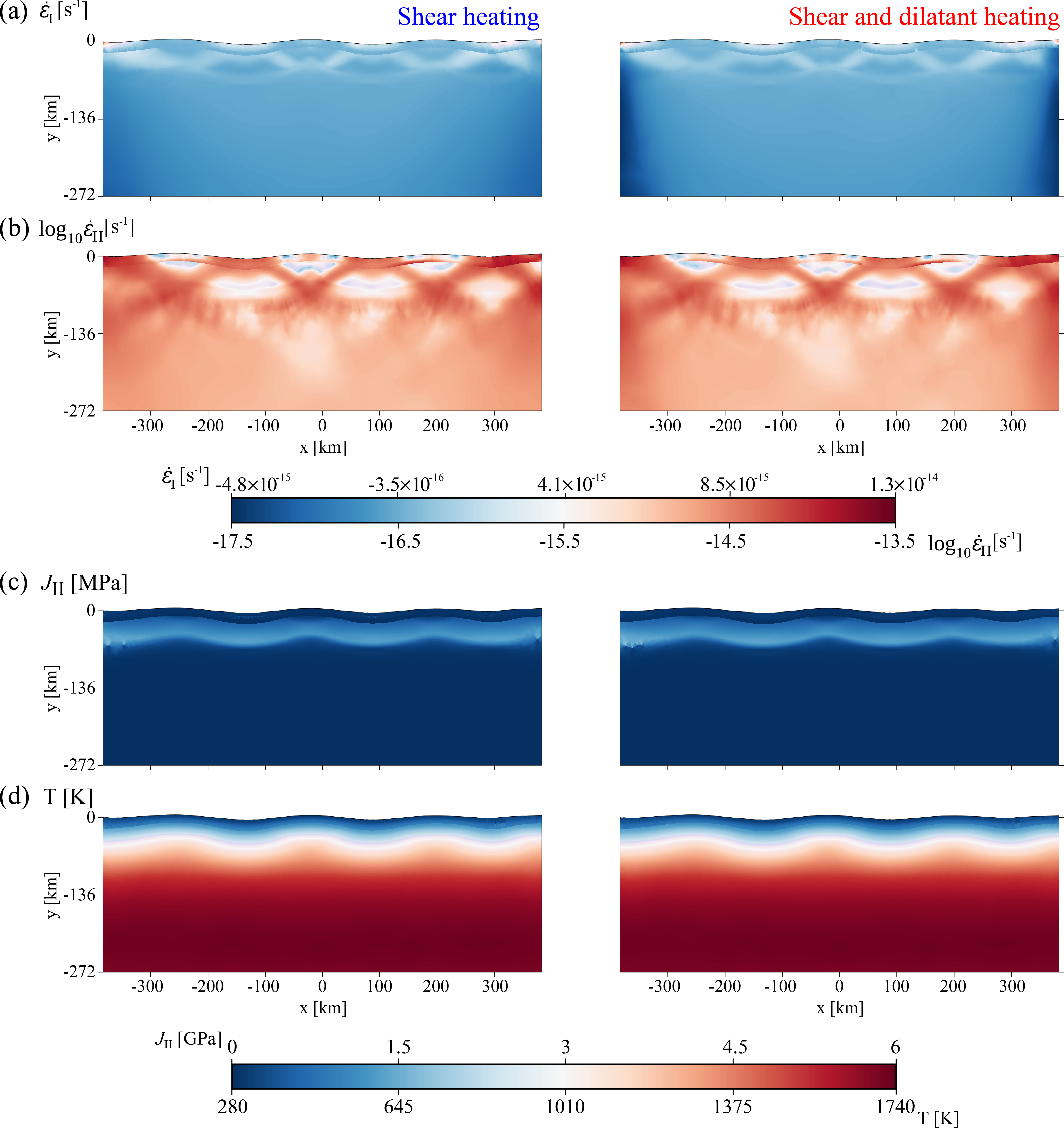}
\caption{Geodynamic model results for setup in Figure \ref{input_geo} without the weak zone after 5.7 million years of horizontal compression corresponding to an overall shortening of 23\%. (a) volumetric strain rate invariant, (b) deviatoric strain rate invariant, (c) second invariant of stress, and (d) evolved temperature.}
\label{buckling_results}
\end{figure}

\subsection{Potential contribution of volumetric plastic deformation to lithospheric localization, and heating}

\subsubsection{The buckling case}

The results of the experiment without the weak zone after 5.7 million years of horizontal compression are shown in Figure \ref{buckling_results}. Seven shear bands developed whether shear or shear and dilatant heating were accounted for (Figure \ref{buckling_results}a-b). The interaction of these shear bands led to lithospheric buckling (Figure \ref{buckling_results}c) and a zone of localization near the center of the model (between x = 0 and -100 km). We also observed that volumetric and shear localization were more enhanced at the center of the model when shear heating was considered alone, i.e., the tent-shaped feature resulting from the intersection of the two central shear bands was stronger when shear heating was the only dissipation mechanism (Figure \ref{buckling_results}a-b). The temperature showed an uplift of isotherms where the shear bands interacted (Figure \ref{buckling_results}d).

\begin{figure}
\centering
\includegraphics[width=11cm]{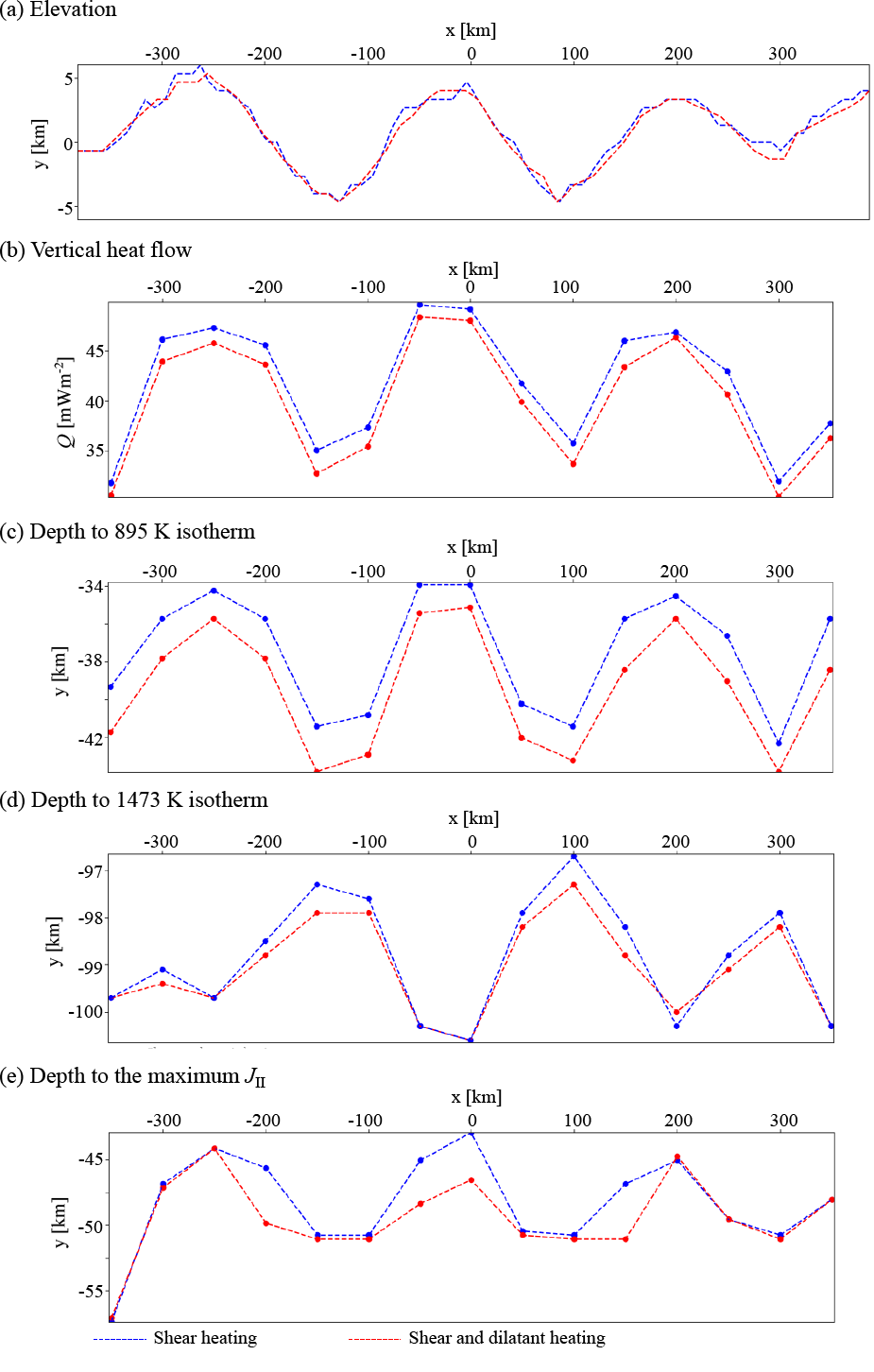}
\caption{Comparative profiles for geodynamic results shown in Figure \ref{buckling_results} (case 1) showing: (a) surface elevation, (b) magnitude of vertical heat flow, (c) depth to 895 K isotherm (d) depth to 1473 K isotherm, and (e) depth to the maximum deviatoric stress where the model shifts from brittle to ductile behaviour.}
\label{buckling_results_extracted_info}
\end{figure}
\begin{figure}
\centering
\includegraphics[width=14cm]{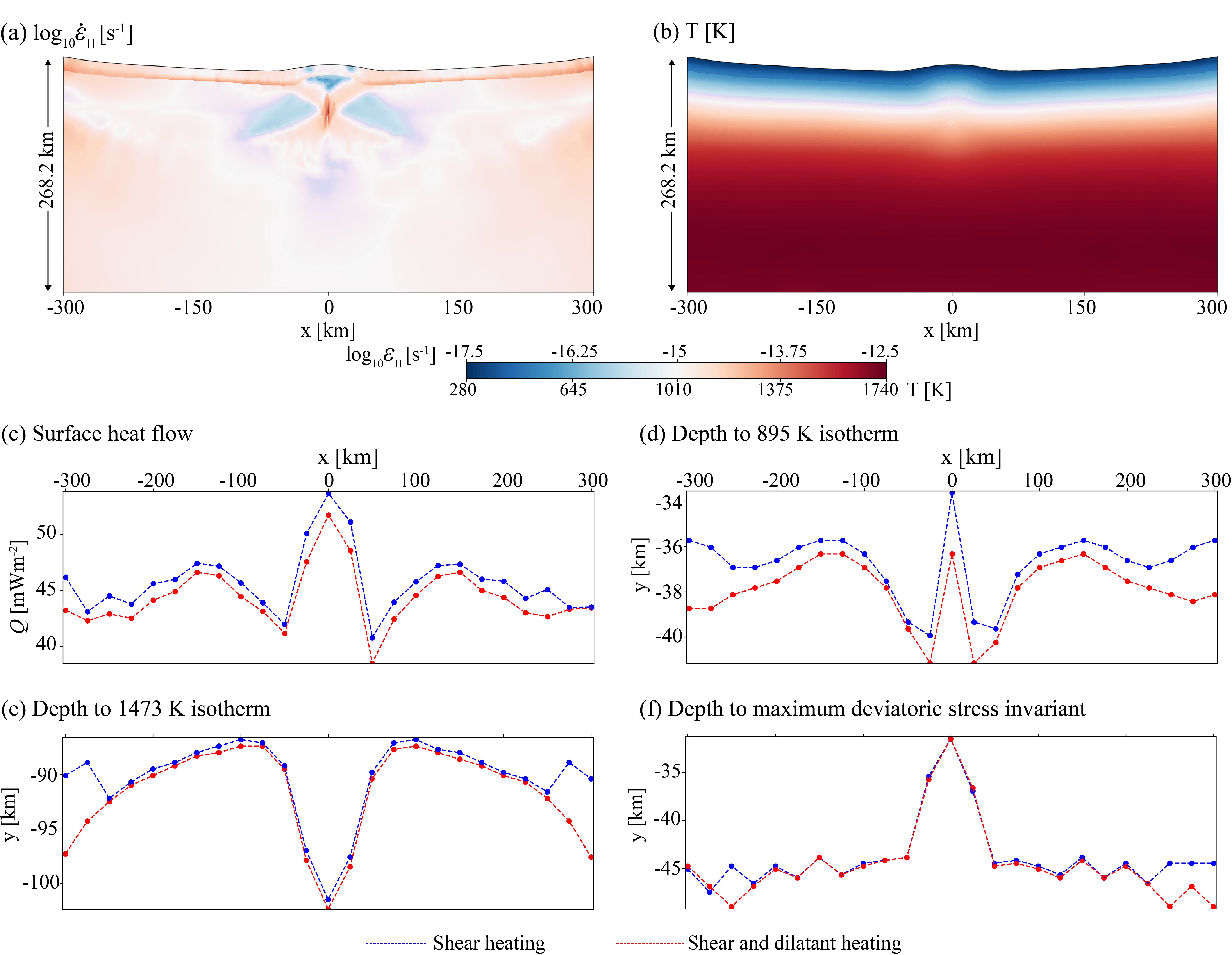}
\caption{Results for input model shown in Figure \ref{input_geo} for shear and dilatant heating (case 2). (a) Logarithm of the deviatoric strain rate invariant, (b) temperature, (c) magnitude of vertical heat flow, (d) depth to 895 K isotherm chosen based on one of the isotherms that was shallower when dilatant heating was included; (e) depth to 1473 K where ductile behaviour was active, and (f) depth of the maximum deviatoric stress invariant which can be used as the transition to brittle mantle behaviour to ductile mantle behaviour.}
\label{central_elliprical_heterogeneity_EII}
\end{figure}
We compared the impact of shear dissipation and shear with volumetric plastic dissipation on the results of the buckling experiment. Figure \ref{buckling_results_extracted_info}a displays the surface elevation for both cases with shear heating and with shear and dilatant heating. We extracted 1-D plots every 50 km along the model to investigate the variations of the magnitude of the vertical heat flow (Figure \ref{buckling_results_extracted_info}b), depth to 895 K isotherm (Figure \ref{buckling_results_extracted_info}c), depth to the 1473 K isotherm (Figure \ref{buckling_results_extracted_info}d) and the depth to the maximum value of the second invariant of the stress tensor where the model switches from brittle behaviour to ductile behaviour (Figure \ref{buckling_results_extracted_info}e)}. The heat flow was calculated based on the temperature within the first 10 km using the standard formulation of the product of thermal conductivity and thermal gradient \cite{england2010}. From the heat flow plots, we observed variations of up to 3 mW s\textsuperscript{-2} reduction in heat flow when dilatant heating was included (Figure \ref{buckling_results_extracted_info}b). We chose the 895 K isotherm as it was observed to be the isotherm that was elevated when shear heating was the dissipation mechanism (Figure \ref{buckling_results}d). This isotherm was found to be offset by up to 2.5 km when volumetric plastic dissipation is included with shear dissipation (Figure \ref{buckling_results_extracted_info}c). The depth to the 1473 K (ensuring viscous flow) indicates that volumetric plastic dissipation influences this isotherm (Figure \ref{buckling_results_extracted_info}d). This suggests that including volumetric plastic dissipation thickens the brittle part of lithosphere by a few kilometers. Finally, the depth of the brittle-ductile transition is also influenced when the dilatant heating is included with shear heating (Figure \ref{buckling_results_extracted_info}e). Considering that the maximum of the deviatoric stress invariant was observed to correspond to the depth of transition from brittle to ductile behaviour (Figure \ref{duretz2021resultsdilatancy1D}b), the results from the buckling experiment shown in Figure \ref{buckling_results_extracted_info}e indicate that the depth to the brittle-ductile transition can be thickened by an additional 3.75 km near x = 0 when dilatant heating is included. 

\subsubsection{The domal case}
Our experimental setup in Figure \ref{input_geo} includes an elliptical weak zone which could represent a low-velocity zone associated with partial melting, internal fluid-filled fractures, high-temperature area or inherited features from a previous geological event. The importance of such weak features in influencing tectonic regimes has been investigated by past authors \cite{cloetingh1982,gurnis1992,nikolaeva2010,ruh2015}. Our interest here is to investigate if we can observe similar features associated with domal structures due to compressional tectonics, for example, an anomalous thermal zone beneath a dome, the formation of faults that may affect the surrounding areas and the surface topography.

For a simulation of 5.4 million years, we show the deformed domain in an area of size 268.2 km by 600 km around the weak inclusion and away from the inevitable edge effects. Furthermore, since the structural evolution between cases accounting for shear heating and shear with dilatant heating were similar, we only show the results of shear with dilatant heating for deformation (Figure \ref{central_elliprical_heterogeneity_EII}a) and temperature (Figure \ref{central_elliprical_heterogeneity_EII}b), where 4 shear bands emanated from the elliptical inclusion. Two of the shear bands propagated upwards to the brittle lithosphere and led to an elevated surface topography, while the other two propagated downwards into the ductile part before fading out (Figure \ref{central_elliprical_heterogeneity_EII}a). This behaviour led to the shallowing and deepening of some isotherms. We however compare results from shear heating and shear with dilatant heating in extracted profile plots which we show in Figures (Figure \ref{central_elliprical_heterogeneity_EII}c-f). The large-scale shear bands and the associated thermal feedback led to a zone with initially horizontal isotherms rising beneath the weak zone. While the dome is associated with elevated surface heat flow for shear or shear and dilatant heating (Figure \ref{central_elliprical_heterogeneity_EII}c), the depth to the 895 K isotherm exhibited a shallowing beneath the dome (Figure \ref{central_elliprical_heterogeneity_EII}d) and a deepening of the 1473 K isotherm (Figure \ref{central_elliprical_heterogeneity_EII}e). This suggests a thermal anomaly which is a consequence of shear and dilatant dissipation. The specific effect of dilatant heating is to shallow the 895 K isotherm by about 2 km beneath the dome. The depth to the maximum of the deviatoric stress invariant used as the point of transition from brittle mantle behaviour to ductile mantle behaviour is unaffected by shear heating or shear with dilatant heating (Figure \ref{central_elliprical_heterogeneity_EII}f). Comparing the results of the depths to the chosen isotherms and the depth to the brittle-ductile transition, we can conclude that the effect of shear heating and shear with dilatant heating mostly influences the brittle domain in this case.  

\section{Discussion}

In this contribution, we have described a formulation for viscoelastic viscoplastic (VE-VP) constitutive behaviour, incorporating elastic compressibility, plastic compressibility, and power law viscoplasticity. We addressed the potential feedback from volumetric plastic dissipation, which we referred to as dilatant heating, and had been traditionally neglected in conventional geodynamic modelling. Our approach therefore treated the domain as a solid-mechanical system which also differs from traditional approaches that utilize a fluid-mechanical idealization of a solid system. We have given a step-by-step description of the constitutive update for stresses and strains; consistent Jacobian matrices for adaptive time-stepping as well as a concise algorithmic description. In this concise discussion section, we shall examine the impact of this volumetric plastic feedback to the dissipative heating loop.

The justification for the effect of volumetric plastic deformation on heat reduction can be first explained in the following  way: volumetric plastic deformation implies a work term whose sign depends ultimately on the constitutive laws. When thermal conductivity is low as is the case for rocks, the overall process should be close to isentropic, which implies that dilatant plasticity (expansion) indeed implies some cooling. From equation \ref{pressure_update}, once the viscoplastic multiplier is determined, a correction to the pressure is implemented. Since the pressure multiplier is always positive, there is a reduction in pressure. 

For pure shear experiments (TEST 2), the impact of volumetric plastic and shear dissipation compared to shear dissipation alone did not show a large difference in the volumetric and deviatoric strain rates; pressure and deviatoric stress invariant. Despite the volume-preserving pure shear experiment, we observed a temperature reduction of up to 3 K when dilatant plastic heating was included. This is because shearing dominated the deformation, and the feedback to volumetric plastic deformations and resultant dilatant plastic heating was minimized, i.e., there was little volume change, despite extensive internal shear deformations. This was indeed apparent in Figure \ref{duretz20results}c and \ref{duretz20results}d, where the different dissipative feedback did not significantly impact the pressure. 

TEST 4A to investigate viscoplasticity, which included an initially isothermal medium, indicated that volumetric plastic dissipation impacted the volumetric strain rate, deviatoric strain rate, pressure, deviatoric stress invariant, and irreversible strains. History-dependent variables like the integrated irreversible strains were also influenced by the contribution of volumetric plastic dissipation. In all cases, the temperature rose with the feedback on the temperature commencing before the other history-dependent variables like accumulated viscoplastic strains and deviatoric stress invariants. High strain rate experiments (TEST 4B) also showed that temperature can be reduced by up to 25 to 50 K when dilatant plastic heating was considered.

It has been argued that rocks generally deform viscoelastically at low stresses and do not necessarily support elastic stresses over the time scale of a subduction process, for example \cite{schmeling2008benchmark}. A metric for assessing the time scale over which elastic stresses can be relaxed is the Maxwell relaxation time, given as the ratio of the effective viscosity and the shear modulus. In some numerical studies, the impact of elasticity is inhibited by setting an unrealistically high value for the shear modulus, which makes the Maxwell relaxation time small, hence elastic stresses are dissipated quickly \cite{schmeling2008benchmark}. We have not included assumptions in our model to mitigate elasticity, as it had been shown in other studies that elasticity (represented by the shear modulus) is an important attribute in lithospheric-scale strain localization \cite{jaquet2016,bessat2020}. Our inclusion of elasticity in our model is justified on the basis that, we assume that the material can deform elastically until thermally-activated viscous flow kicks in, while viscoplastic behaviour evolves once the Drucker-Prager yield criterion is breached. 

In utilizing our new implementation in studying lithospheric dynamics, our results indicated that realistic heat flow values were reproducible for thermal dissipation alone when compared with published results of heat flow measurements from compressional environments  \cite{manga2012,ezenwaka2023}. We also observed that the contribution of volumetric plastic dissipation either for the buckling experiment or for the collisional case was a reduction of heat flow values, and variation in the depth of some isotherms, with shear heating more likely to lead to lithospheric thinning than shear with dilatant plastic heating.

As observed in all the simulations reported here, it is much more common for several shear bands to develop (see Figures \ref{buckling_results}a and \ref{buckling_results}b, for example). This sets up the question of what happens where different shear bands intersect/interact. In these regions, deformation is both intense and going in all directions which impedes the development of a single dominant shear direction. Hence localization in the sense of allowing a simple macroscopic shear direction to develop is hampered. Nevertheless, deformational heating remains intense because strong irreversible deformation is still occurring in such regions (Figure \ref{buckling_results}d).

\section{Conclusion}

We have developed a solid-thermomechanical constitutive description to study strain localization in compressional experiments with contributions from volumetric plastic dissipative feedback, which are usually not considered in geodynamic models. Using small strain assumptions, we developed and applied a full 2-D viscoelastic-viscoplastic stress update scheme and included time adaptivity with consistent algorithmic tangent moduli. Our constitutive laws have been implemented on the Abaqus\textsuperscript{\textcopyright} finite element solver. Based on our studies, we conclude as follows:
\setlist[enumerate]{label={\Roman*.}}
\begin{enumerate}
\item We benchmarked our new solid-mechanical model against other published numerical models for viscoelastic and viscoelastic-viscoplastic rheologies. Our benchmarking tests indicate that our simulations are not mesh-sensitive nor are they sensitive to low-order (linear) or high-order (quadratic) elements.
\item In comparing results of shear heating and shear with dilatant heating for pure shear numerical experiments, our results indicated that the implicit conservation of volume in the experimental setup through the specific choice of boundary conditions to keep the background strain rate constant leads to the volumetric plastic feedback not having a strong influence on the results. 
\item A departure from volume-conserving boundary conditions pure shear experiments demonstrated the contributions of volumetric plastic feedback to the temperature and the stress state. Using sequentially a linear temperature profile (TEST 3),  an isothermal case that ensured viscoplastic deformation (TEST 4A), and a high boundary strain rate (TEST 4B), we observed that volumetric plastic dissipation contributed to the thermal state of our models by reducing the amount of heat produced, which affected the depth to isotherms and even the brittle-ductile transition. 
\item Whether thermal dissipation is due to shear heating or shear with dilatant heating, both scenarios led to a temperature rise, but the inclusion of volumetric plastic dissipation led to reduced heating, sometimes up to 8\%.
\item The volumetric plastic dissipation leading to reduced heating is primarily due to plastic dilation (expansions). Despite including dilatant plasticity for the mantle, our results did not indicate that volumetric plastic dissipation had significant effects in the hotter parts of the mantle as we did not include assumptions of volumetric plastic strains during ductile creep. Our new model is therefore able to switch between viscoelastic behaviour (controlled by the temperature) and viscoplastic behaviour (controlled by the Drucker-Prager yield criterion which occurs in the brittle regimes).
\item We also successfully applied our new model to lithospheric-scale dynamics where we found that whether shear heating and shear with dilatant heating are accounted for, the onset of buckling develops at the same time. However, where we observed a higher volumetric plastic strain, we observed a temperature reduction of 28 - 33 K, which is about 4\% reduction of shear heating when dilatant terms are included.
\end{enumerate}

While the actual temperature change is not so large, we have shown that the impact of accounting for volumetric plastic stresses and strains on several aspects of crustal-scale and lithosperic-scale geodynamic models is not negligible, such as the depths to isotherms, brittle-ductile transition; deformation history, stress evolution, and heat flow.  Our new model also allowed the emergence of several complexities which were not imposed a priori, for example, softening behaviour. The fact that the brittle-ductile transition is a function of the amount of plastic dilatancy also opens new vistas to other questions, for example, the influence on mantle and crustal solidus, thermal steady-state i.e., where the amount of heat produced is evacuated by conduction thereby keeping the temperature in equilibrium; and investigating areas with high heat flows that can be targeted for geothermal exploration.

\section*{Acknowledgements}
This work was supported by G\'{e}oscience Environment Toulouse/Observatoire Midi-Pyr\'{e}n\'{e}es, Institut Physique du Globe de Paris (IPGP), INTERREG V Cara\"{i}bes program, the European Regional Development Fund (FEDER through the PREST project and the AXA Research Fund. EM and ST acknowledge the funding support. HSB appreciates the European Research Council grant PERSISMO (grant number 865411) for partially supporting this work. Dr Nadaya Cubas and Professor Laetitia Le Pourhiet are greatly appreciated for their support and insights; as well as Professor Thibault Duretz for offering useful suggestions which enhanced the quality of the paper. Special appreciation to Antoine Jacquey who thoroughly reviewed an earlier version of this manuscript, and one anonymous reviewer who motivated us to benchmark our new model. Finally, we thank Dr D\`aniel Kiss, the two anonymous reviewers of the revised manuscript, the assistant editor, Dr Louise Alexander, and the editor, Dr Tobias Keller, who copiously reviewed this work. This is IPGP contribution number: ....
\section*{Data availability}
All codes developed in the framework of the work and input files to reproduce the experiments reported here are stored in \url{https://github.com/ekemomoh007/VEVP-Code-and-Input-Files}.

\bsp 
The appendices in this guide were generated by typing:
\clearpage
\appendix

\section{Numerical Implementation of Constitutive Laws}\label{AppendixA}
The formulations to solve the mechanical and thermal diffusion parts of the system described in Section 2 have been coded into a User-defined Mechanical Material behaviour (UMAT) and User-defined Thermal Material behaviour UMATHT subroutines, respectively, and implemented in Abaqus\textsuperscript{\textcopyright} \cite{abaqustheory,abaqus}. 

Given a displacement at a given time step which can result from a velocity or force-driven boundary condition, Abaqus\textsuperscript{\textcopyright} supplies the total strain increment at that time step. Using our proposed constitutive laws, corresponding elastic stresses are computed. The viscous and viscoplastic corrections to the stresses are at the heart of the stress update scheme, and for an adaptive time-stepping scheme that we adopt, the Consistent Algorithmic Tangent Modulus is required. We describe the details of our algorithmic implementation here.
\subsection{Stress Update Scheme\label{stressupdate1}}
At each time step, we first assume that the strain increment, ${{\Delta{\varepsilon}^\textrm{e}_{ij}}}$, at that time step is fully elastic. We then compute an elastic trial stress.
\begin{equation}
{\sigma}^{\textrm{e, trial}}_{ij} = C^\textrm{e}_{ijkl}\left({\varepsilon}^\textrm{e, trial}_{kl}+{\Delta\varepsilon}^\textrm{e, trial}_{kl}\right)
.\end{equation}
Given the elastic trial stress, we obtain trial deviatoric stresses and pressures from:
\begin{equation} 
\begin{aligned}
{s_{ij}^\textrm{e,trial}=\sigma_{ij}^\textrm{e,trial}-\dfrac{\sigma_{kk}^\textrm{e, trial}}{3}}\delta_{ij},\\  {P^\textrm{e,trial}=\dfrac{I_1^\textrm{e, trial}}{3}=\dfrac{\sigma_{kk}^\textrm{e, trial}}{3}}. 
\end{aligned}
\end{equation}
The trial deviatoric stresses are used to compute the invariant of deviatoric stresses in the elastic state. Given the function in Equation \ref{functional_form}, we can write a viscous function in the form:
\begin{equation}
{\tilde\Phi^\textrm{v}}{({\sigma_{ij}^\textrm{e}},T)=\left({\Phi^\textrm{v}}\right)^mf(T)-\dfrac{\Delta{\gamma}^\textrm{v}}{\Delta{t}}}=0,
\label{viscous_functional_form}
\end{equation}
where ${\Delta{t}}$ is the time increment at a step, ${{\Phi^\textrm{v}}=\sqrt{J_\textrm{II}^\textrm{e}(s_{ij}^\textrm{e, trial})}-G\Delta\gamma^\textrm{v}}$. Equation \ref{viscous_functional_form} is a non-linear equation and we obtain the unknown ${\Delta\gamma^\textrm{v}}$ by performing Newton-Raphson iterations \cite{benisrael66,yqma} using the derivative of ${{\tilde\Phi^\textrm{v}}}$ with respect to ${\Delta\gamma^\textrm{v}}$:
\begin{equation}
\dfrac{{\textrm{d}\tilde\Phi^\textrm{v}}}{{\textrm{d}\Delta\gamma^\textrm{v}}}={=-mG\left({\Phi^\textrm{v}}\right)^{m-1}f(T)\Delta{t}-1}=0,
\label{viscous_functional_form_NR}
\end{equation}  
After computing ${\Delta\gamma^\textrm{v}}$, update the deviatoric stresses as follows:
\begin{equation}
\Delta{\varepsilon}_{ij}^\textrm{v} = \Delta{\gamma^\textrm{v}}({\sigma_{ij}^\textrm{e, trial}},T)\dfrac{\partial{\Phi_\textrm{F}^\textrm{v}}}{\partial{s}_{ij}^\textrm{e,trial}}=\Delta{\gamma^\textrm{v}}\dfrac{{s}_{ij}^\textrm{e,trial}}{2\sqrt{J_\textrm{II}^\textrm{e}(s_{ij}^\textrm{e, trial})}},
\label{viscousflow_viscous_strains}
\end{equation}
\begin{equation}
{{{s}_{ij}^\textrm{v} ={\left(1-\dfrac{G\Delta{\gamma^\textrm{v}}}{\sqrt{J_\textrm{II}^\textrm{e}({s}^{\textrm{e, trial}}_{ij})}}\right)}{s}_{ij}^{\textrm{e, trial}}}}
\label{viscousflow_viscous_dev_stresses}
\end{equation}
A deviatoric stress invariant (${J_\textrm{II}^\textrm{v}}$) is computed after correcting for creep. This new stress invariant is fed to assess the viscoplastic routine through the Drucker-Prager yield function in Equation \ref{drucker}. If the yield criterion is violated, we perform more Newton-Raphson iterations to obtain the viscoplastic multiplier. We therefore write a residual viscoplastic potential function using the functional form of \cite{zien1974,perzyna1986}:
\begin{equation}
\tilde{\Phi}^\textrm{vp}_\textrm{R}{({\sigma_{ij}^\textrm{v}},T)=\dfrac{\Delta{t}}{\mu}\left({\dfrac{\Phi_\textrm{R}^\textrm{DP}}{\Phi_{0}}}\right)^m-\Delta{\gamma}^\textrm{vp}}=0.
\label{residual_functionals_vp}
\end{equation}
Where ${{\Phi_\textrm{R}^\textrm{DP}}= \Phi_\textrm{Y}^\textrm{DP}-c_1\Delta\gamma^\textrm{vp}}$, with ${c_1 = (G+\alpha_1 \alpha_3 K)}$, ${\Phi_\textrm{0}=c_0}$ and ${\mu}$ is a viscosity-related term. Because Equation \ref{residual_functionals_vp} is a non-linear equation, we find the unknown ${\Delta\gamma^\textrm{vp}}$  by Newton-Raphson iterations as the creep case. The derivative of the residual viscoplastic potential function required for these iterations is given by:
\begin{equation}
\dfrac{\textrm{d}{\tilde\Phi_\textrm{R}^\textrm{vp}}}{\textrm{d}\Delta\gamma^\textrm{vp}} = -\dfrac{c_{1}m\Delta{t}}{c_0 \mu}\left(\dfrac{\Phi_\textrm{Y}-c_{1}\Delta\gamma^\textrm{vp}}{c_0}\right)^{m-1} - 1
\label{residual_functionals_vp_derive}
\end{equation}
Equations \ref{residual_functionals_vp} and \ref{residual_functionals_vp_derive} are used to perform Newton-Raphson iterations to obtain the unknown ${\Delta\gamma^\textrm{vp}}$. A critical aspect of the convergence of the Newton-Raphson iterations for obtaining both the viscous and viscoplastic multipliers is the requirement that the multipliers are strictly positive and that the algorithm converges. To aid the convergence of the Newton-Raphson scheme, we have included a modified bisection method when these conditions are violated \cite{net,chapra}. 

After obtaining the viscoplastic multiplier upon yielding, the stresses, viscoplastic strains, temperature change, and state variables can be computed:  
\begin{equation}
\Delta{\varepsilon}^\textrm{vp}_{ij}  = \Delta{\gamma^\textrm{vp}}\left(    \dfrac{{s}^{\textrm{v}}_{ij}}{2\sqrt{J_\textrm{II}^\textrm{v}({s}^{\textrm{v}}_{ij})}}+ \dfrac{\alpha_3}{3}\delta_{ij}\right)
\label{vpupdate}
\end{equation}
\begin{equation}
{{{s}_{ij}^\textrm{vp} ={\left(1-\dfrac{G\Delta{\gamma^\textrm{vp}}}{\sqrt{J_\textrm{II}^\textrm{v}({s}^{\textrm{v}}_{ij})}}\right)}{s}_{ij}^{\textrm{v}}}},
\label{deviatoric_update}
\end{equation}
\begin{equation}
{ P=P^{\textrm{e, trial}}-{\alpha_3}\Delta{\gamma^\textrm{vp}}K},
\label{pressure_update}
\end{equation}
\begin{equation}
\sigma_{ij}=s_{ij}^\textrm{vp}+P,
\label{stress_update_vp}
\end{equation}
\begin{equation}
\Delta{T}=\dfrac{\sigma_{ij}\Delta{\varepsilon_{ij}^\textrm{vp}}}{\rho C_p}.
\label{temperaturechange}
\end{equation}
\subsection{Consistent Algorithmic Tangent Modulus (CATM)\label{stressupdate2}}
Finally, using the Newton scheme to solve the global equations, an important consideration is the convergence of the solver and the selection of an adaptive time step. To this end, we estimate a convergence matrix obtained by linearizing the stress update schemes for deviatoric stresses and pressure given in Equations \ref{viscousflow_viscous_dev_stresses}, \ref{deviatoric_update} and \ref{pressure_update}. We proceed by taking small increments in the deviatoric stress and pressure with respect to the trial elastic state. We compute this matrix for the creep and viscoplastic states.
\subsubsection{Consistent Algorithmic Tangent Modulus for Creep Deformation}

\begin{equation}
C_{ijkl}^\textrm{v}={\dfrac{\textrm{d}{\sigma}_{ij}}{\textrm{d}\varepsilon_{kl}^{\textrm{e, trial}}}=\dfrac{\textrm{d}{s}_{ij}^\textrm{v}}{\textrm{d}\varepsilon_{kl}^{\textrm{e, trial}}}+\delta_{ij}\left(\dfrac{\textrm{d}P}{\textrm{d}\varepsilon_{kl}^{\textrm{e},\;\textrm{trial}}}\right)}
\end{equation}
The expression for deviatoric stresses can be equivalently written in terms of the deviatoric strains (${{e}_{ij}^{\textrm{e, trial}}}$) taking  into account that ${s_{ij}^\textrm{e, trial}=2G{e}_{ij}^{\textrm{e, trial}}}$, and ${{e}_{ij}^{\textrm{trial}}}={{\varepsilon}_{ij}^{\textrm{trial}}}-{{\varepsilon}_{kk}^{\textrm{trial}}}/3$:
\begin{equation}
{{{s}_{ij}^\textrm{v} =2G{\left(1-\dfrac{\Delta{\gamma}^\textrm{v}}{\sqrt{2}{{e}_{\textrm{norm}}^{\textrm{trial}}}}\right)}{e}_{ij}^{\textrm{trial}}}}.
\end{equation}
${{e}_{\textrm{norm}}^{\textrm{trial}}= \sqrt{{e}_{ij}^{\textrm{trial}}{e}_{ij}^{\textrm{trial}}}}$ represents the Euclidian norm of the deviatoric strain tensor. 
The elements to assemble the consistent Jacobian matrix during creep deformation are given by derivatives of the stress states after creep: 

\begin{eqnarray}
\dfrac{\textrm{d}{s}_{ij}^\textrm{v}}{\textrm{d}\varepsilon_{kl}^{\textrm{e}\;\textrm{trial}}}  = 2G{\left(1-\dfrac{\Delta{\gamma^\textrm{v}}}{\sqrt{2}{e}_{\textrm{norm}}^{\textrm{trial}}}\right)}\mitbf{\upi}^{\textrm {dev}}_{ijkl}\nonumber\\+{\sqrt{2}G}\left(\dfrac{\Delta{\gamma^\textrm{v}}}{e_\textrm{norm}^{\textrm{trial}}}-b_1\right)\dfrac{{e}_{ij}^{\textrm{trial}}}{{e}_{\textrm{norm}}^{\textrm{trial}}}\dfrac{{e}_{kl}^{\textrm{trial}}}{{e}_{\textrm{norm}}^{\textrm{trial}}},
\label{deviatoric contribution}
\end{eqnarray}
{\centering{
\begin{eqnarray}
\delta_{ij}\dfrac{\textrm{d}P^\textrm{e, trial}}{\textrm{d}\varepsilon_{kl}^{\textrm{e, trial}}}&=K\delta_{ij}\delta_{kl},
\label{volumetric contribution}
\end{eqnarray}}}
with
\begin{equation*}
b_1=\dfrac{\sqrt{2}G\left(\Delta tf(T)\right)^{1/m}}{\dfrac{{(\Delta\gamma^\textrm{v}})^{(1-m)/m}}{m}+G(\Delta
 t f(T))^{1/m}}
\end{equation*}
\noindent The expression for the consistent Jacobian, when creep deformation is considered without viscoplasticity, is given by a combination of Equations {\ref{deviatoric contribution}} and \ref{volumetric contribution}:
\begin{eqnarray}
C_{ijkl}^\textrm{v}=2G{\left(1-\dfrac{\Delta{\gamma^\textrm{v}}}{\sqrt{2}{e}_{\textrm{norm}}^{\textrm{trial}}}\right)}\mitbf{\upi}^{\textrm {dev}}_{ijkl} = \nonumber \\+\sqrt{2}G\left(\dfrac{\Delta{\gamma^\textrm{v}}}{e_\textrm{norm}^{\textrm{trial}}}-b_1\right)\dfrac{{e}_{ij}^{\textrm{trial}}}{{e}_{\textrm{norm}}^{\textrm{trial}}}\dfrac{{e}_{kl}^{\textrm{trial}}}{{e}_{\textrm{norm}}^{\textrm{trial}}} + K\delta_{ij}\delta_{kl}
\end{eqnarray}
\noindent where ${\mitbf{\upi}_{ijkl}^{\textrm{dev}}  = \left(\delta_{ik}\delta_{jl} + \delta_{il}\delta_{jk}\right)/2-\delta_{ij}\delta_{kl}/3}$ is the fourth order deviatoric projection tensor, which extracts the deviatoric components of a stress or strain tensor.
\subsubsection{Consistent Algorithmic Tangent Modulus for Viscoplastic Deformation}
The linearization scheme follows the same procedure used for the creep state. For the deviatoric stresses,
\begin{equation}
{{\dfrac{\textrm{d}{s}_{ij}^\textrm{vp}}{\textrm{d}\varepsilon_{kl}^{\textrm{e}\;\textrm{trial}}} ={\left(1-\dfrac{G\Delta{\gamma}^\textrm{vp}}{\sqrt{J_\textrm{II}^{\textrm{v}} ({s}_{\textrm{ij}}^{\textrm{v}})}}\right)}}}\dfrac{\textrm{d}{s}_{\textrm{ij}}^{\textrm{v}}}{\textrm{d}\varepsilon_{kl}^{\textrm{e}\;\textrm{trial}}}.
\end{equation}

\begin{eqnarray*}
{{\dfrac{\textrm{d}{s}_{ij}^\textrm{vp}}{\textrm{d}\varepsilon_{kl}^{\textrm{e}\;\textrm{trial}}} ={\left(1-\dfrac{G\Delta{\gamma}^\textrm{vp}}{\sqrt{J_\textrm{II}^{\textrm{v}} ({s}_{\textrm{ij}}^{\textrm{v}})}}\right)}}}\dfrac{\textrm{d}{s}_{\textrm{ij}}^{\textrm{v}}}{\textrm{d}\varepsilon_{kl}^{\textrm{e}\;\textrm{trial}}} -\dfrac{\alpha_1b_2KG}{\sqrt{J_\textrm{II}^\textrm{v}({s}_{\textrm{ij}}^{\textrm{v}})}}\delta_{ij}\dfrac{s_{kl}^\textrm{v}}{\sqrt{J_\textrm{II}^\textrm{v}(s_{kl}^\textrm{v})}}-\dfrac{G}{\sqrt{J_\textrm{II}^\textrm{v}(s_{ij}^\textrm{v})}}\\\left(\sqrt{2}Gb_2\left(1-\dfrac{b_1}{\sqrt{2}}\right)\dfrac{e_{ij}^\textrm{trial}}{e_\textrm{norm}}+\dfrac{\sqrt{2}G\Delta\gamma^\textrm{vp}}{\sqrt{J_\textrm{II}^\textrm{v}(s_{ij}^\textrm{v})}} \left(\dfrac{b_1}{\sqrt{2}}-1\right)\dfrac{e_{ij}^\textrm{trial}}{e_\textrm{norm}}\right)\dfrac{s_{kl}^\textrm{v}}{\sqrt{J_\textrm{II}^\textrm{v}(s_{kl}^\textrm{v})}},
\label{deviatoric contribution_vp}
\end{eqnarray*}
{{
\begin{eqnarray}
\delta_{ij}\dfrac{\textrm{d}P_{n+1}}{\textrm{d}\varepsilon_{kl}^{\textrm{e}\;\textrm{trial}}}=K\left(1-{\alpha_1\alpha_3b_2K}\right)\delta_{ij}\delta_{kl}+{\alpha_3b_2K}\left(b_1-\sqrt{2}G\right)\dfrac{{e}_{ij}^{\textrm{trial}}}{{e}_{\textrm{norm}}^{\textrm{trial}}}\delta_{kl},
\label{volumetric contribution_vp}
\end{eqnarray}}}
where 
\begin{equation*}
b_2=\dfrac{\dfrac{1}{c_0}\left(\dfrac{\Delta t}{\mu}\right)^{1/m}}{\dfrac{\left(\Delta\gamma^\textrm{vp}\right)^{(1-m)/m}}{m}+\dfrac{c_1}{c_0}\left(\dfrac{\Delta t}{\mu}\right)^{1/m}}.
\end{equation*}

The procedure of obtaining the tangent moduli is a linearization of the stress update scheme, such that the adaptive time stepping is a function of the deformation state of the material. 

\end{document}